\documentclass[aps,nopacs,showkeys,nofootinbib,preprint]{revtex4}
\usepackage{epsfig,amssymb,bm,graphicx,color,amsmath,rotating}
\usepackage{xcolor}
\usepackage[percent]{overpic}
\usepackage{mwe,xcolor}
\usepackage{transparent}
\usepackage{amsmath}
\usepackage{framed}
\usepackage{slashed}
\usepackage{diagbox}
\usepackage[export]{adjustbox}
\usepackage{sidecap}
\usepackage{hyperref}
\newcommand{\arX}[1]{\href{https://#1}{doi:#1}.}
\newcommand\arx[3][r]{
  \ifx r#1 \href{https://arxiv.org/abs/#2}{[arXiv:#2].} \else
  \ifx o#1 \href{https://arxiv.org/abs/#3/#2}{[arXiv:#3/#2].} \else
  \ifx b#1 \href{https://arxiv.org/abs/#2}{[arXiv:#2 [#3]].} \else
  {Illegal~option}
  \fi\fi\fi
}

\begin{document} {\normalsize} 

\title{Towards a warm holographic equation of state by an Einstein-Maxwell-dilaton model} 

\author{R.~Z\"ollner${}^1$, B.~K\"ampfer${}^{2, \, 3}$}
\affiliation{${}^1$Institut f\"ur Technische Logistik und Arbeitssysteme, TU~Dresden, 01062 Dresden, Germany}
\affiliation{${}^2$Helmholtz-Zentrum  Dresden-Rossendorf, 01314 Dresden, Germany}
\affiliation{${}^3$Institut f\"ur Theoretische Physik, TU~Dresden, 01062 Dresden, Germany}
 
\begin{abstract}

The holographic Einstein-Maxwell-dilaton model is employed to map state-of-the-art lattice QCD thermodynamics data
from the temperature ($T$) axis towards the baryon-chemical potential ($\mu_B$)
axis aimed at gaining a warm equation of state (EoS) of deconfined QCD matter 
which can be supplemented with a cool and confined part 
suitable for subsequent compact (neutron) star (merger) investigations.
The model exhibits a critical end point (CEP) at
$T_\mathrm{CEP} = \mathcal{O}(100)$~MeV and $\mu_{B \, \mathrm{CEP}} = 500 \ldots 700$~MeV
with emerging first-order phase transition (FOPT) curve which extends to large values of $\mu_B$
without approaching the $\mu_B$ axis.
We consider the impact and peculiarities of the related phase structure on the EoS for the
employed dilaton potential and dynamical coupling parameterizations.
These seem to prevent to design an overall trustable EoS without recourse to hybrid constructions.

\end{abstract}


\date{\today}

\maketitle

\section{Introduction} \label{sect:introduction}

The advent of detecting gravitational waves of merging neutron stars \cite{LIGOScientific:2020aai}
and improved determinations of mass-radius relations of neutron stars, 
e.g.\ by NICER \cite{Miller:2019cac,Riley:2019yda,Miller:2021qha},
triggered a firework of related investigations,
most notably focused on the access to the cool equation of state (EoS) of dense strong-interaction matter
on the theory side.   
Among the various approaches to compact (neutron) star EoS is the application of the famous AdS/CFT correspondence
which mimics dense matter by a suitable gravity dual of QCD \cite{Jarvinen:2021jbd,Hoyos:2021uff,Chesler:2019osn}.
Here, we employ such a holographic approach based on the Einstein-Maxwell-dilaton (EMd) model 
pioneered in \cite{DeWolfe:2010he,DeWolfe:2011ts} and used further on in 
\cite{Grefa:2022sav,Zhang:2022uin,Cai:2022omk,Critelli:2017oub,Grefa:2021qvt,Knaute:2017opk}.
References \cite{Rougemont:2023gfz} and \cite{Jokela:2024xgz} provide 
a survey and a valuable comparison with Dirac-Born-Infeld AdS/CFT models. 

Our motivation is as follows. Given lattice QCD thermodynamics results, e.g.\ the scaled pressure,
$p/T^4$, on the temperature ($T$) axis \cite{Borsanyi:2021sxv,HotQCD:2014kol} and supplementing them
by the susceptibility $\chi_2 = \partial^2 p / \partial \mu_B^2\vert_{\mu_B=0}$ \cite{Borsanyi:2013bia},
the EMd model delivers primarily entropy density $s(T, \mu_B)$ and baryon density $n_B(T, \mu_B)$ 
which can be integrated to arrive at the potential $p(T, \mu_B)$.\footnote{
Even at $\mu_B = 0$, $p(T, \mu_B =0) = p(T_{min}) + \int_{T_{min}}^T \mathrm{d} \bar T s(\bar T)$
is a non-local quantity which needs $p(T_{min})$ as input parameter, supposed $s(T \ge T_{min}, \mu_B = 0)$
is accurately given. More favorable is to use directly $p(T, \mu_B =0)$ from lattice QCD results
and map it without further assumptions into the $T$-$\mu_B$ plane, as advocated below.
}   
These quantities depend on the baryon-chemical potential $\mu_B$ as well. 
The lattice input \cite{Borsanyi:2021sxv} is provided, at $\mu_B = 0$, for $T \in [T_1, T_2] = [125, 240]$~MeV 
and allows to adjust the dilaton potential and the dynamical coupling of the EMd model.
The lattice data \cite{Borsanyi:2021sxv} for  
$\mu_B/T = 0.5, 1, \cdots, 3.5$ 
serve as control, again in $T \in [125, 240]$~MeV as save for 2+1 flavors.
(For other relevant lattice data sets, cf.\ \cite{Bazavov:2017dus,Borsanyi:2022qlh,Bollweg:2022fqq}.)
Curves of $p(T, \mu_B) = const$, i.e.\ isobars, can be utilized then to continue the lattice-given EoS
$p(T, \mu_B =0)$ into the $T$-$\mu_B$ plane and, finally -- if no obstacle is met --, to $p(T=0, \mu_B)$.
That is, the EoS -- here the pressure -- is directly mapped from the $T$ axis on the $\mu_B$ axis.
Curves of $p(T, \mu_B) = const$ are determined by solving
\begin{align} \label{eq:T(muB)}
\left. \frac{\mathrm{d} T (\mu_B)}{\mathrm{d} \mu_B} \right\vert_{p=const} 
= - \frac{n_B (T, \mu_B)}{s(T, \mu_B)} ,
\end{align}
where the inverse entropy-per-baryon determines the slope field.
Solutions are $T(\mu_B)\vert_{p=p_0}$ with $T(\mu_B=0)\vert_{p=p_0} = T_0$
and $p(T_0, \mu_B =0) = p_0$.
Via Gibbs-Duhem, the energy density ($e$) follows from $e = - p + s T + n_B \mu_B$.

The gained cool EoS $p(e)$ can be then used as input for compact (neutron) star calculations.
This vision is illustrated in Fig.~\ref{fig:Fig1} in Appendix \ref{sect:appD} with a toy model. 
The above mentioned state-of-the-art lattice data uncovers in fact a relevant pressure interval:
$p(T_1, \mu_B = 0) = \mathcal{O}(10^1)~\textrm{MeV/fm}^3$ and 
$p(T_2, \mu_B = 0) = \mathcal{O}(10^3)~\textrm{MeV/fm}^3$.
For the toy model, these values translate into
$p(T = 0, \mu_{B \, 1}) = \mathcal{O}(10^1)~\textrm{MeV/fm}^3$ and 
$p(T = 0, \mu_{B \, 2}) = \mathcal{O}(10^3)~\textrm{MeV/fm}^3$
localized just above the reliable pressure interval accessible by nuclear-physics many-body methods
(cf.\ figure 1 in \cite{Ecker:2022dlg,Annala:2021gom} and figure 12 in \cite{Hebeler:2013nza}
which suggest the matching point at about $3$~MeV/fm${}^3$ (pressure) and $200$~MeV/fm${}^3$  (energy density)). 
The corresponding energy densities 
$e(T = 0, \mu_{B \, 1})$ and $e(T = 0, \mu_{B \, 2})$ depend of course on the details of the mapping
along $p = const$ curves from the $T$ axis to the $\mu_B$ axis,
as the actual values of $\mu_{B \, 1,2}$.
  
However, such a vision meets potential obstacles.
These are (i) already for $\mu_B = 0$, an unwanted first-order or Hawking-Page phase transition
may occur outside the controlled region $T \in [T_1, T_2]$, where $T_{1, 2}$ are again the
lower and upper limits of the save EoS, which is used to adjust dilaton potential and dynamical coupling
of the EMd model,
(ii) a critical end point (CEP) and related first-order phase transition (FOPT) curve may disturb the expected pattern of curves
$T (\mu_B)\vert_{p = const}$, displayed in Appendix \ref{sect:appD} for the toy model, 
and
(iii) the expected pattern of curves $T (\mu_B)\vert_{p = const}$ does not continue smoothly to the
$\mu_B$ axis. In fact, item (i) is met in \cite{Knaute:2017opk} ($T_\textrm{FOPT} (\mu_B = 0) \approx 45$~MeV),
which however focuses on the neighborhood of a CEP at $\mu_B > 0$, 
and in \cite{Grefa:2021qvt}  ($T_\textrm{FOPT} (\mu_B = 0) \approx 1$~MeV).
The occurrence of a CEP at $\mu_B > 0$ is known \cite{DeWolfe:2010he,DeWolfe:2011ts} and a welcome effect
to study, within such a model class, its impact on the EoS and related dynamics, thus explicating the conjecture
posed in \cite{Stephanov:1999zu,Karsch:2001cy,Fukushima:2010bq,Halasz:1998qr} 
and motivating a continuation of the ongoing energy scan at RHIC \cite{Almaalol:2022xwv,Du:2024wjm}
(For the tight connection of heavy--ion-neutron--star physics, cf.\ \cite{Lovato:2022vgq,MUSES:2023hyz}.) 

To avoid irritations by item (i), we optionally impose here novel side conditions for the dilaton potential.
It happens that this enforces parameterizations which facilitate a turn of the FOPT curve below the CEP from convex
to concave, i.e.\ the FOPT curve levels off and seems to asymptote to the $\mu_B$ axis, thus hindering a concise
cool EoS prediction. Nevertheless, the warm EoS, characterized by $p(e)\vert_T$ 
or the scaled trace anomaly/conformality measure $\Delta (e)\vert_T := (e - 3 p)/(3 e)\vert_T$ 
\cite{Marczenko:2022jhl,Fujimoto:2022ohj} is accessible and may be useful for numerical studies. 

While many preceding EMD studies focus on the CEP localization 
and/or transport coefficients near the CEP and across the FOPT curve, 
we put here emphasize on the EoS. 
Among the hitherto less discussed issues is the pattern of isentropes near the CEP and FOPT.
In \cite{Knaute:2017opk}, it has been pointed out that the isentropes, w.r.t.\ ``incoming" and ``outgoing"
under adiabatic expansion, in various EMd models can be fairly different. 
A general classification of such isentropic patterns has been proposed recently in \cite{Pradeep:2024cca}.
Such a study and holographic perspective is of relevance for heavy-ion collisions experiments and 
speculations on sourcing stochastic gravitation waves in the cosmic confinement transition 
\cite{He:2023ado,Middeldorf-Wygas:2020glx}.

Our paper is organized as follows. We recap the EMd model and its data adjustment in Section~\ref{sect:EMd}.
Numerical results are presented in Section \ref{sect:EoS}, where we discuss CEP position, FOPT curve,
the shape of the $p=const$ curves and the resulting EoS
by means of various contour plots and related cross sections of thermodynamic quantities. 
We conclude and summarize in Section \ref{sect:conclusions}. 
A brief series of appendices complements our analysis.
Appendix~\ref{sect:appD} sketches our vision of mapping QCD data from the $T$ axis into the
$T$-$\mu_B$ plane, and
Appendix~\ref{sect:appA} supplements details of the EMd model used.
Appendix~\ref{sect:appC} considers the density and pressure near and across the FOPT curve.
Appendix~\ref{sect:appB} presents a numerical study of a particular dilaton potential parameterization adjusted to data. 

\section{Holographic Einstein-Maxwell-dilaton model}\label{sect:EMd}

In line with \cite{DeWolfe:2010he,Grefa:2021qvt,Critelli:2017oub} 
we employ the EMd action\footnote{Reference \cite{Hoyos:2021uff}
relates the EMd model (\ref{eq:EMd_action}) to a D3-D7 action and Taylor expanded versions
of string theory-anchored approaches. Hadron, nucleon and quark degrees of freedom are
added separately in \cite{Zhang:2022uin}.}  
in a fiducial five-dimensional pseudo-Riemann space-time with asymptotic AdS symmetry:
\begin{equation}
S_\textrm{EMd} = \frac{1}{2 \kappa_5^2} \int \textrm{d}^4 x \, \textrm{d} r \, \sqrt{-g_5}
\left( R - \frac12 \partial^M \phi \, \partial_M \phi - V(\phi) 
 - \frac14 {\cal G} (\phi) F_{{\cal B}}^2 \right) , 
\label{eq:EMd_action}
\end{equation}
where $R$ is the Einstein-Hilbert gravity part,
$F_{\cal B}^{MN} = \partial^M {\cal B}^N - \partial^N {\cal B}^M$
stands for the field strength tensor of an Abelian gauge field ${\cal B}$ \`{a} la Maxwell
with ${\cal B}_M \textrm{d} x^M = \Phi (r) \, \textrm{d} t$  defining the electro-static potential.
An embedded black hole facilitates the description of a hot and dense medium, since the black hole has
a Hawking surface temperature and sources an electric field, thus encoding holographically 
a temperature and a density of the system.
Dynamical objects are a dilatonic (scalar) field $\phi$ and a Maxwell-type field $\Phi$
which are governed by a dilaton potential
$V(\phi)$ and a dynamical coupling ${\cal G}(\phi)$ and geometry-related quantities.
Space-time is required to be described by the line element squared
\begin{align} \label{eq:metric}
\mathrm{d}s^2=g_{MN}\mathrm{d}x^M\mathrm{d}x^N
= e^{2A(r)}
\left( -f(r) \mathrm{d}t^2+\mathrm{d}\vec{x \, }^2 \right) 
+ \frac{\mathrm{d}r^2}{f(r)} ,
\end{align}
where $r = 0$ is the horizon position, $r \in [0 , \infty]$ the radial coordinate, 
$A$ the warp factor and $f$ the blackness function.
The resulting Einstein equations are a set of coupled second-order ODEs to be solved with appropriate boundary conditions,
see Appendix \ref{sect:appA}.
Within the present bottom-up approach,
the quantities $V$ and ${\cal G}$ are tuned\footnote{``Tuning" within the bottom-up approach faces three issues: 
choices of the functional forms of $V$ and $\mathcal{G}$ and parameter adjustments at data.
For Bayesian analyses cf.\ \cite{Hippert:2023bel}.
}
to reproduce the lattice QCD data  \cite{Borsanyi:2021sxv,Bellwied:2015lba}.

Our ans\"atze for the dilaton potential $V$ and dynamical coupling $\mathcal{G}$ are
\begin{align} \mathcal{W} \equiv
\partial_\phi \ln V(\phi) &= (p_1 \phi + p_2 \phi^2 + p_3 \phi^3 ) \exp\{ - \gamma \phi\} , \label{eq:V}\\
{\cal G} (\phi) &= \frac{1}{1 + c_3 } \left( \frac{1}{\cosh (c_1 \phi + c_2 \phi^2)}
+ \frac{c_3}{\cosh c_4 \phi} \right) \label{eq:F}
\end{align}  
with parameters $\{ p_{1,2,3} \} = \{ 0.165919, $ $0.269459, $ $-0.017133 \}$, 
$ \gamma = 0.471384$
and $\{ c_{1, 2, 3, 4}\} = \{-0.276851, $ $0.394100, $ $0.651725, $ $101.6378 \}$;
the pre-factor $1/(1+c_3)$ ensures $\mathcal{G} (\phi=0) = 1$.
The scale is set by $L^{-1} =  216$~MeV,
in $L^2 V(\phi) = - 12 \exp\{ \int_0^{\phi} \mathrm{d} \tilde \varphi \mathcal{W} (\tilde \varphi) \}$ 
and $\kappa_5 = 1.87 L^{3/2}$ is used.  
The conditions $ \lim_{\phi \to \infty} \mathcal{W}(p_3 > 0) = 0$ and monotony of $V$ towards the boundary 
exclude a purely thermal phase transition at $\mu_B = 0$, cf.\ \cite{Yaresko:2015ysa,Zollner:2018uep}.
The here chosen parameters facilitate a smooth shape of $\mathcal{W}$ with maximum of $0.605$
at $\phi = 3.373$ and a zero at $\phi = 16.321$ which corresponds to an exeedingly small temperature
far below 1~MeV.
Appendix~\ref{sect:appB} presents more details on the impact of the dilaton potential. 

In contrast to former work, we here put emphasis on the optional side condition $p_3 > 0$
which ensures that, at $\mu_B = 0$,
no phase transition is facilitated outside the temperature range uncovered by the lattice data.
Details of our handling of the EMd model are relegated to the Appendix \ref{sect:appA}.

\begin{figure}[h!]
\hspace*{0.6cm } $p/T^4$ \hspace*{2.7cm} $n_B/T^3$ \hspace*{3cm} $e/p$ \\
\includegraphics[width=0.25\columnwidth]{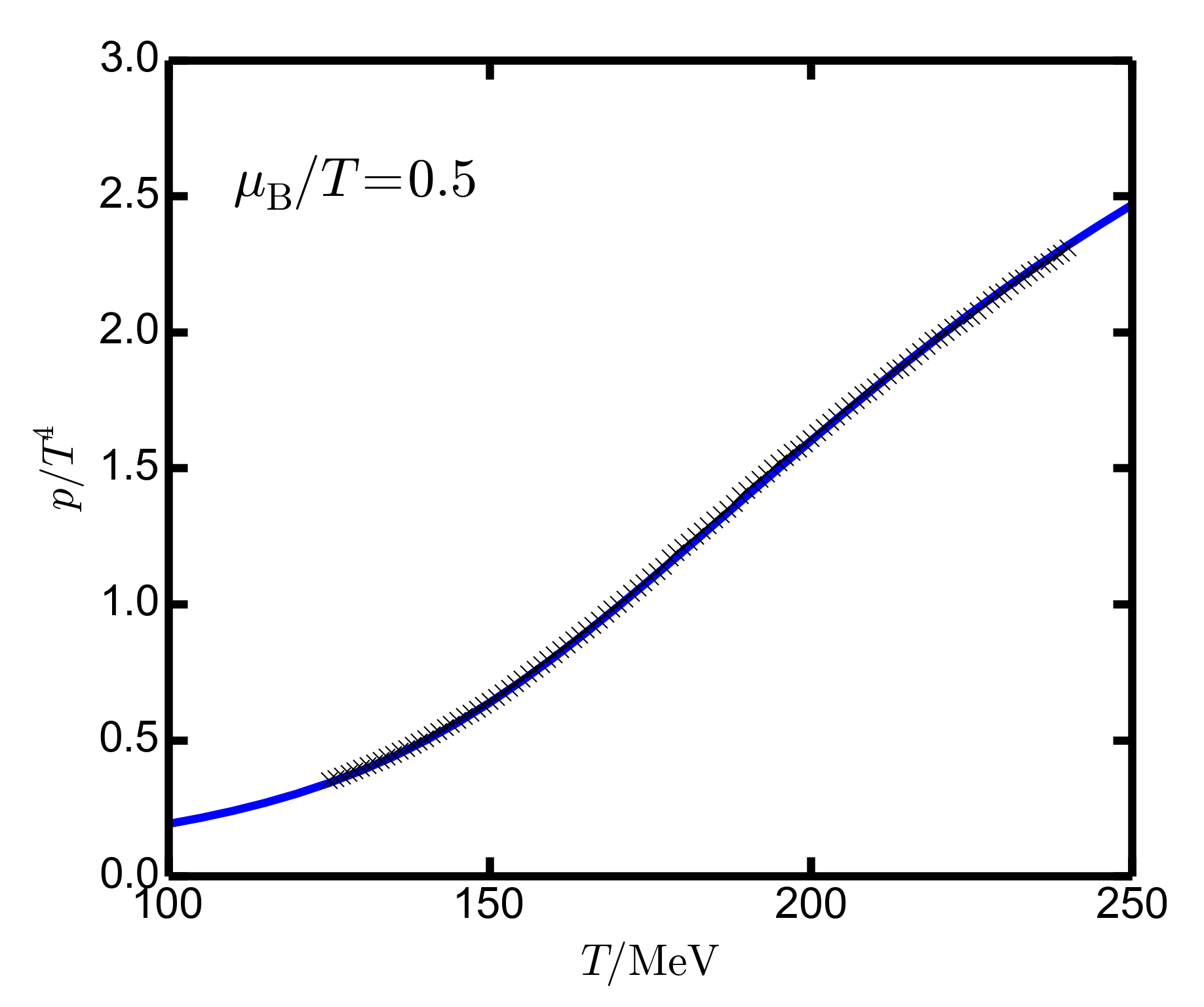}
\includegraphics[width=0.25\columnwidth]{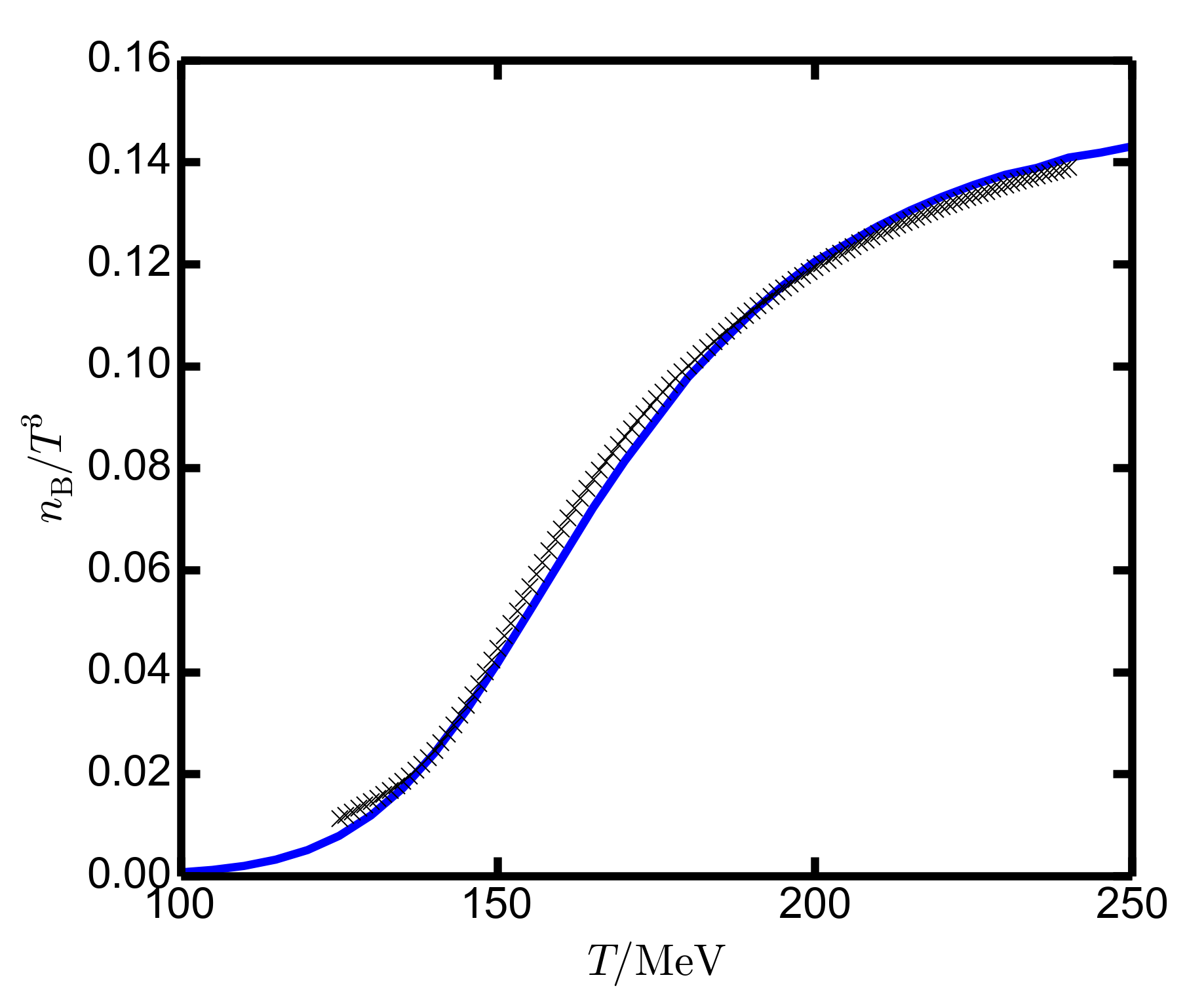}  
\includegraphics[width=0.25\columnwidth]{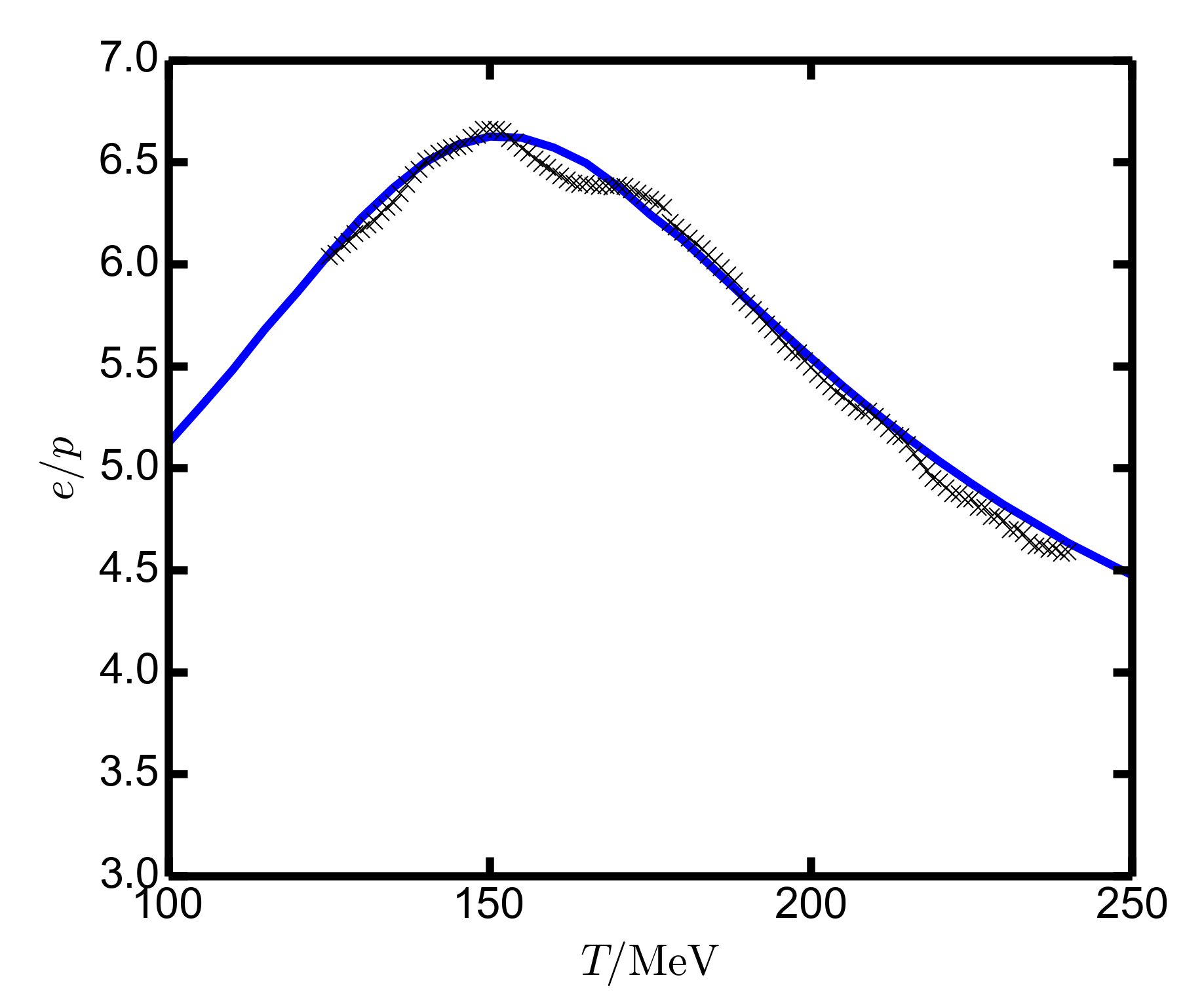}   \\
\includegraphics[width=0.25\columnwidth]{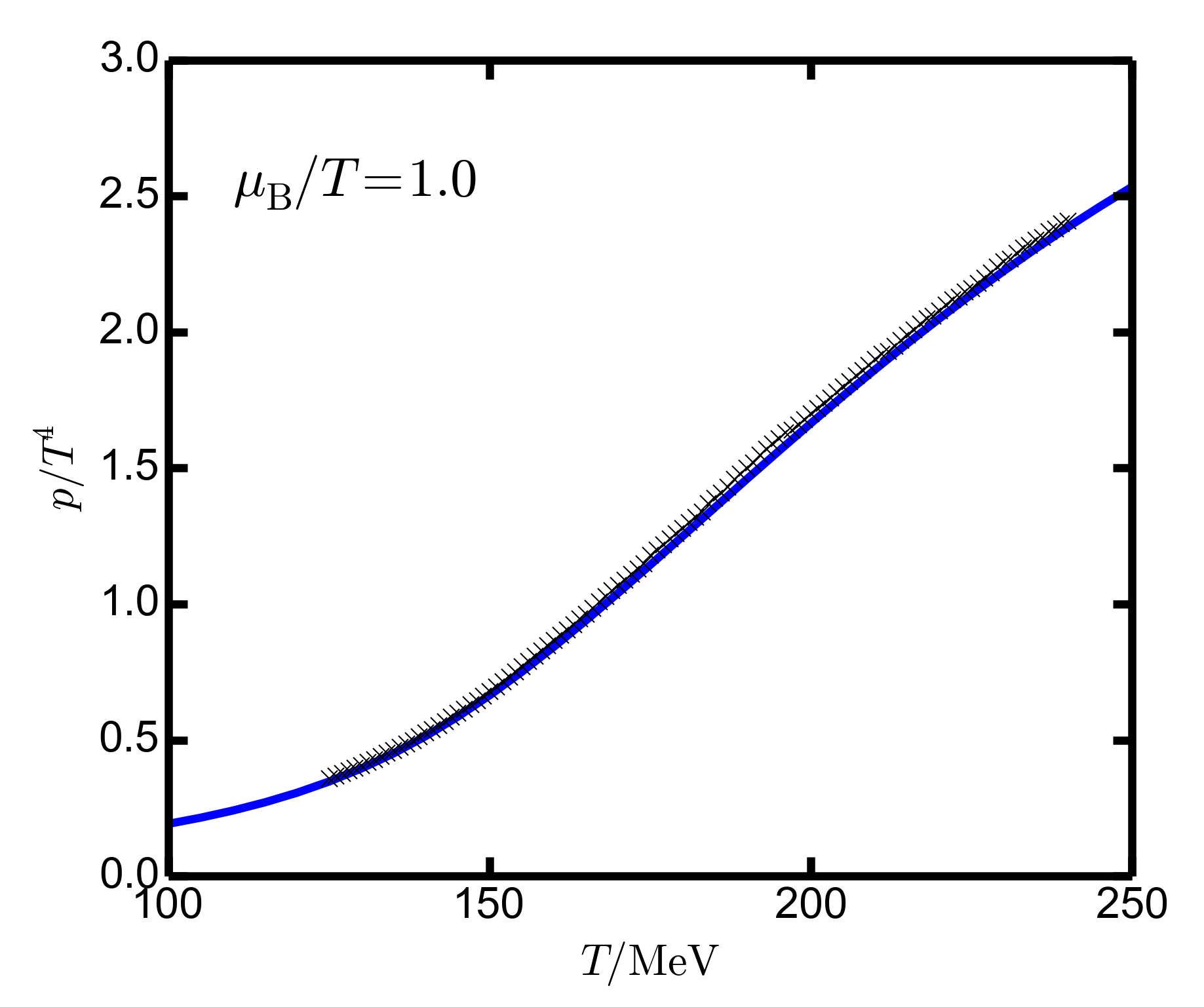}
\includegraphics[width=0.25\columnwidth]{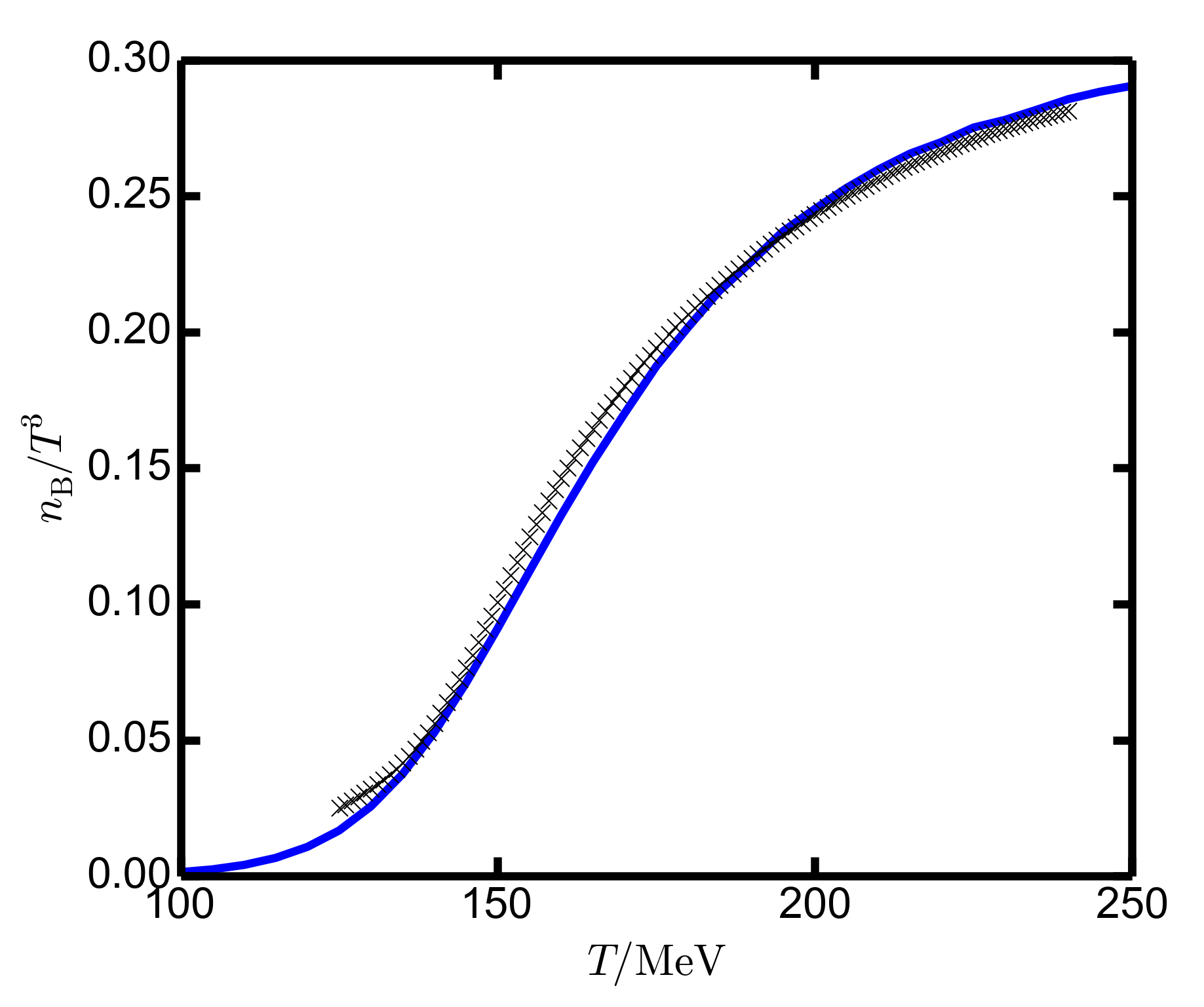}  
\includegraphics[width=0.25\columnwidth]{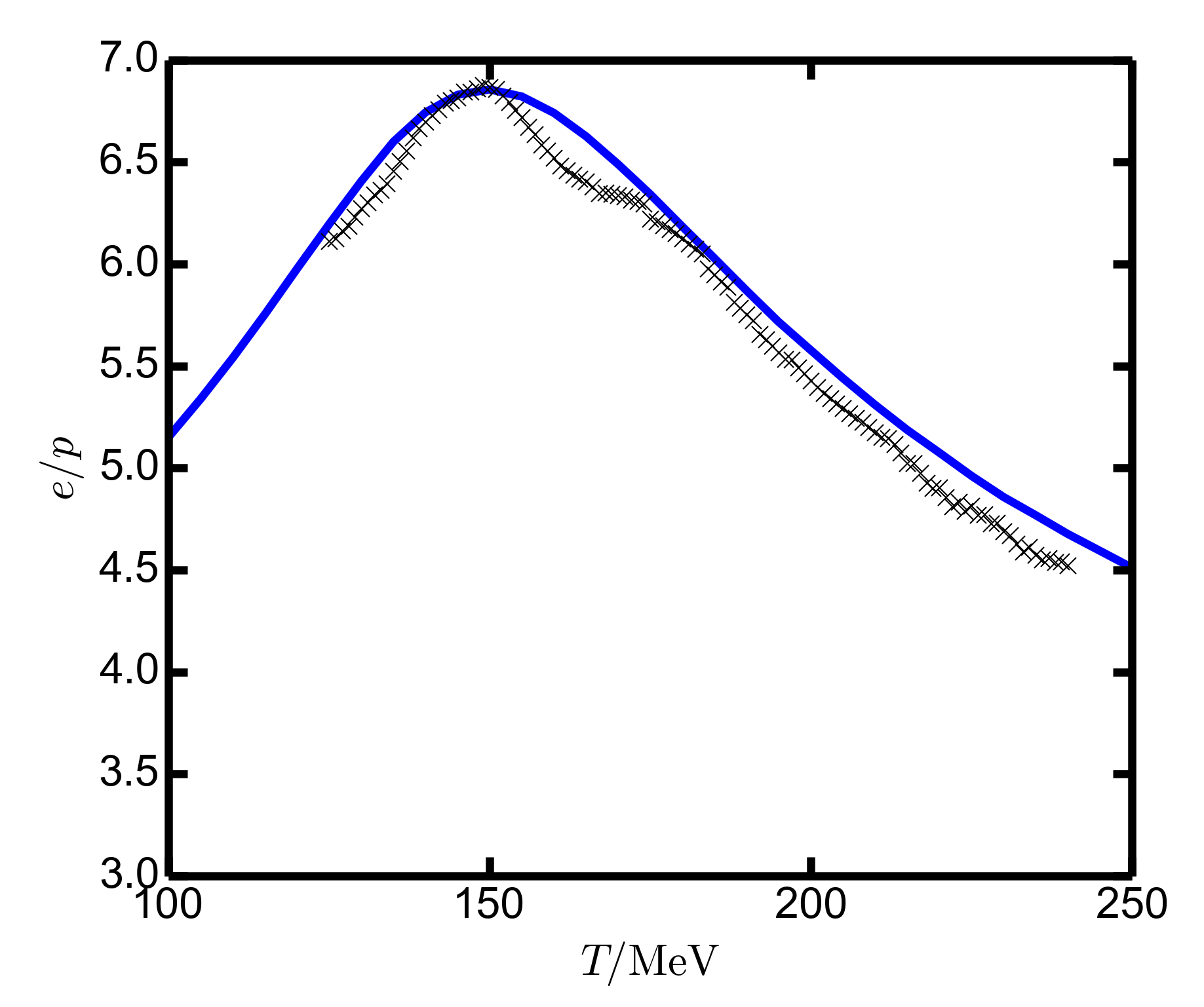} \\
\includegraphics[width=0.25\columnwidth]{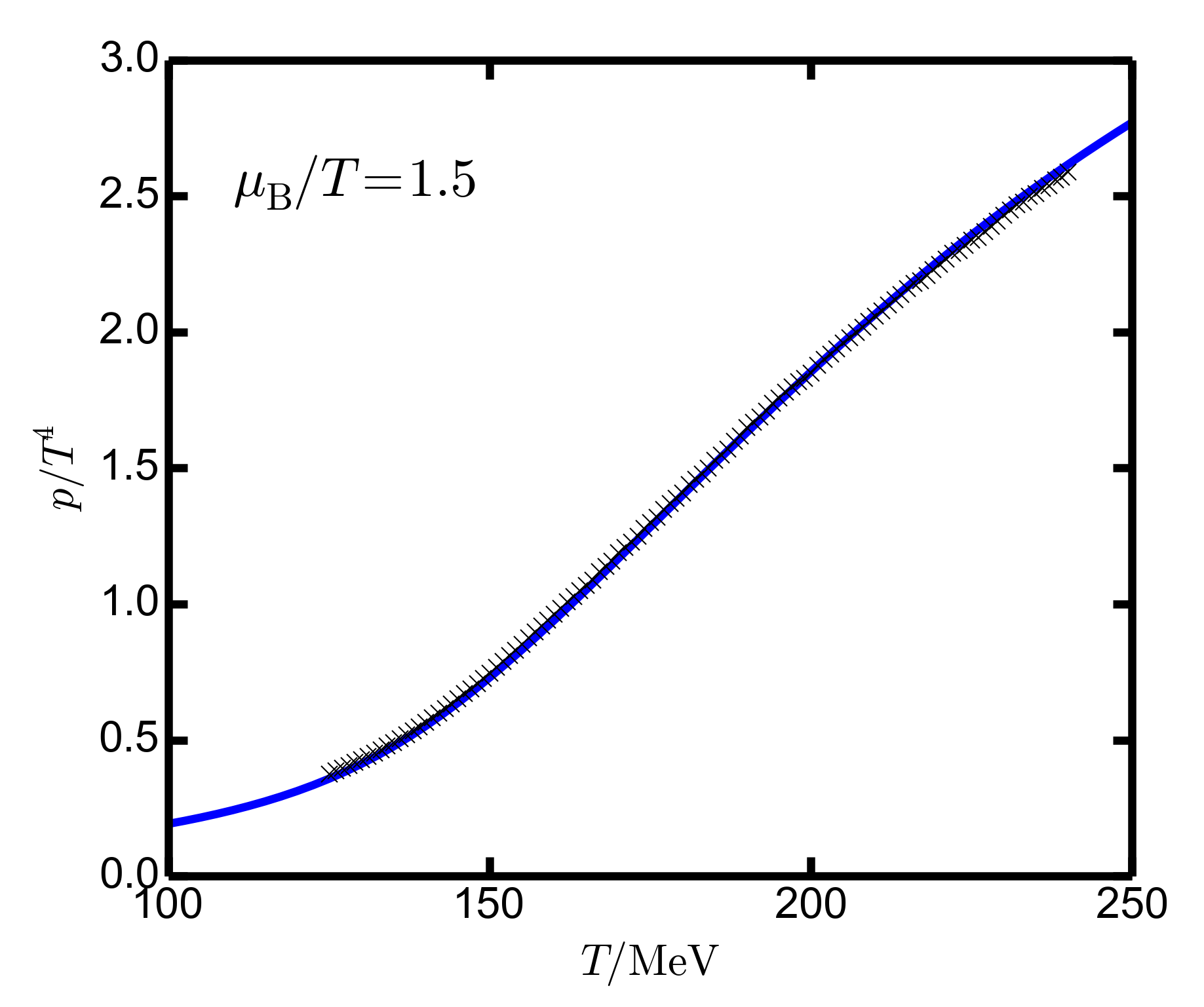}
\includegraphics[width=0.25\columnwidth]{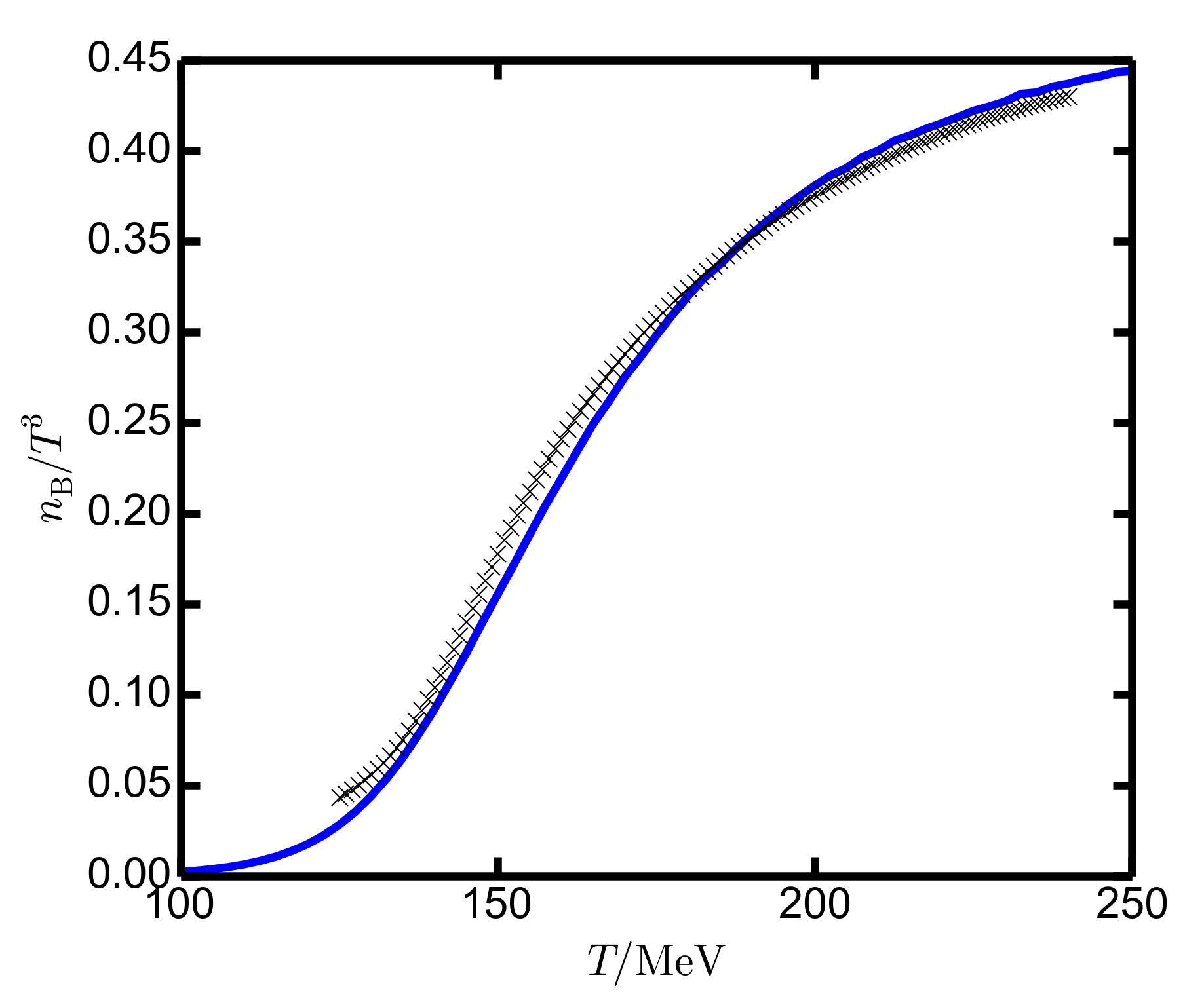}  
\includegraphics[width=0.25\columnwidth]{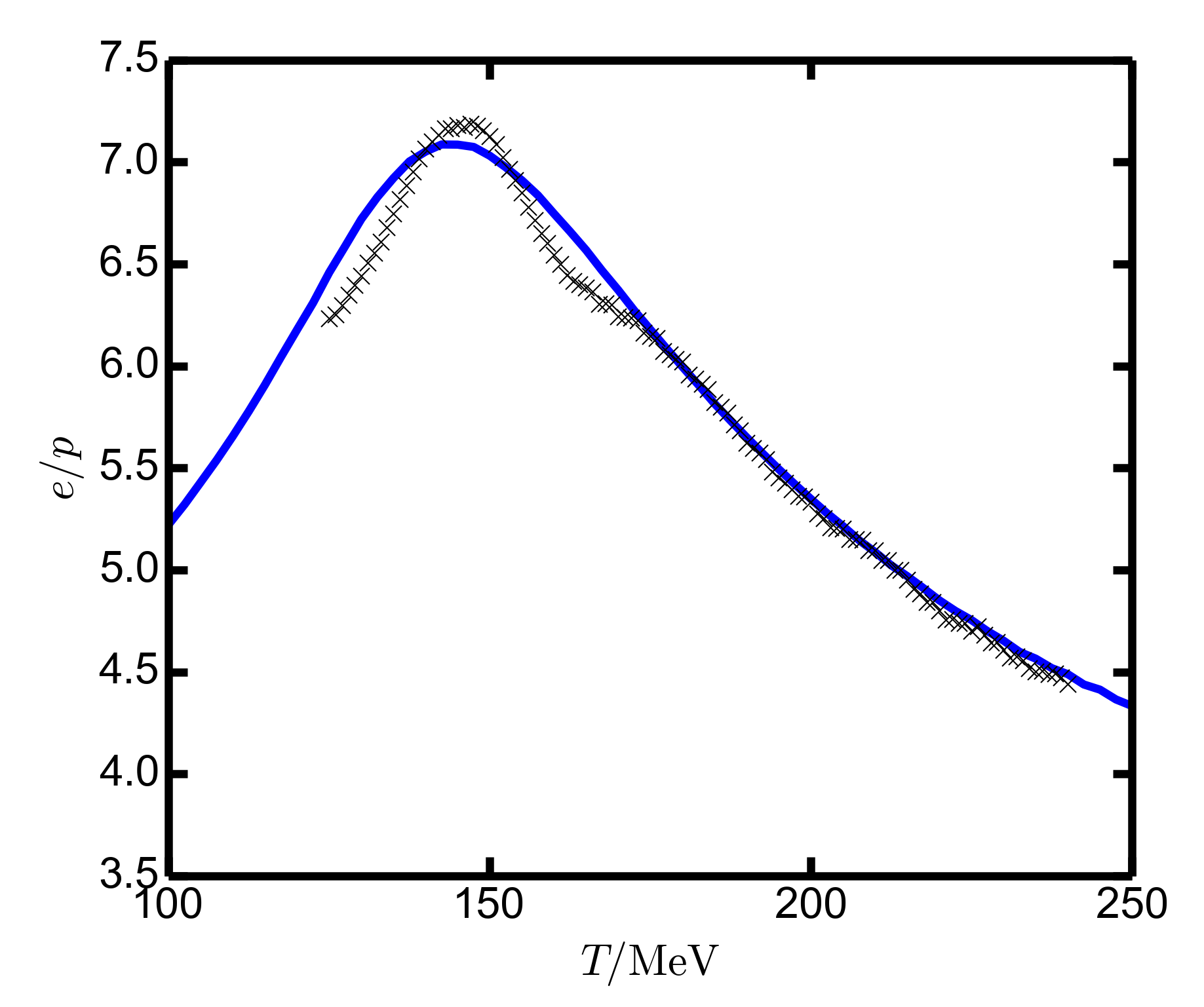}   \\
\includegraphics[width=0.25\columnwidth]{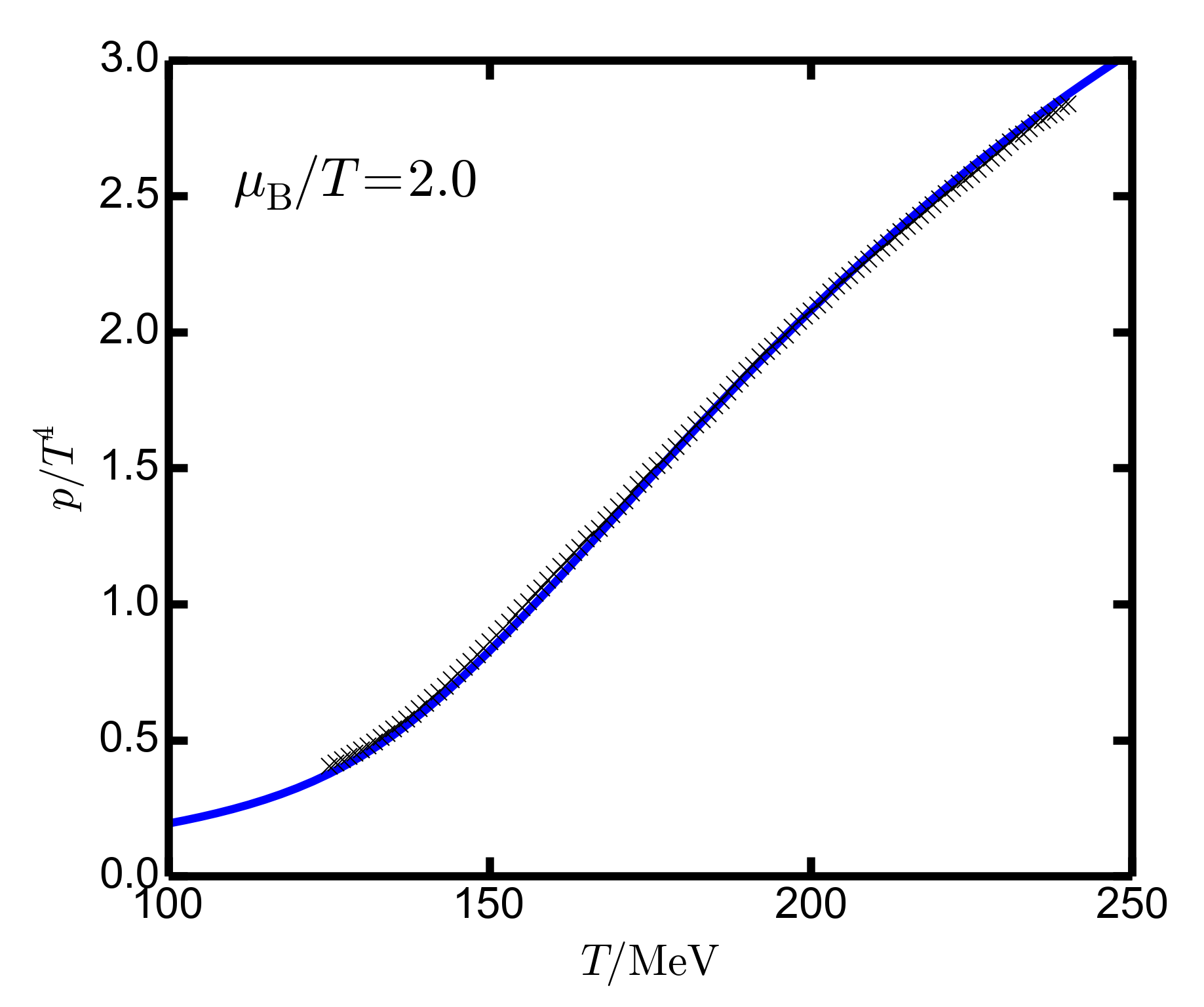}
\includegraphics[width=0.25\columnwidth]{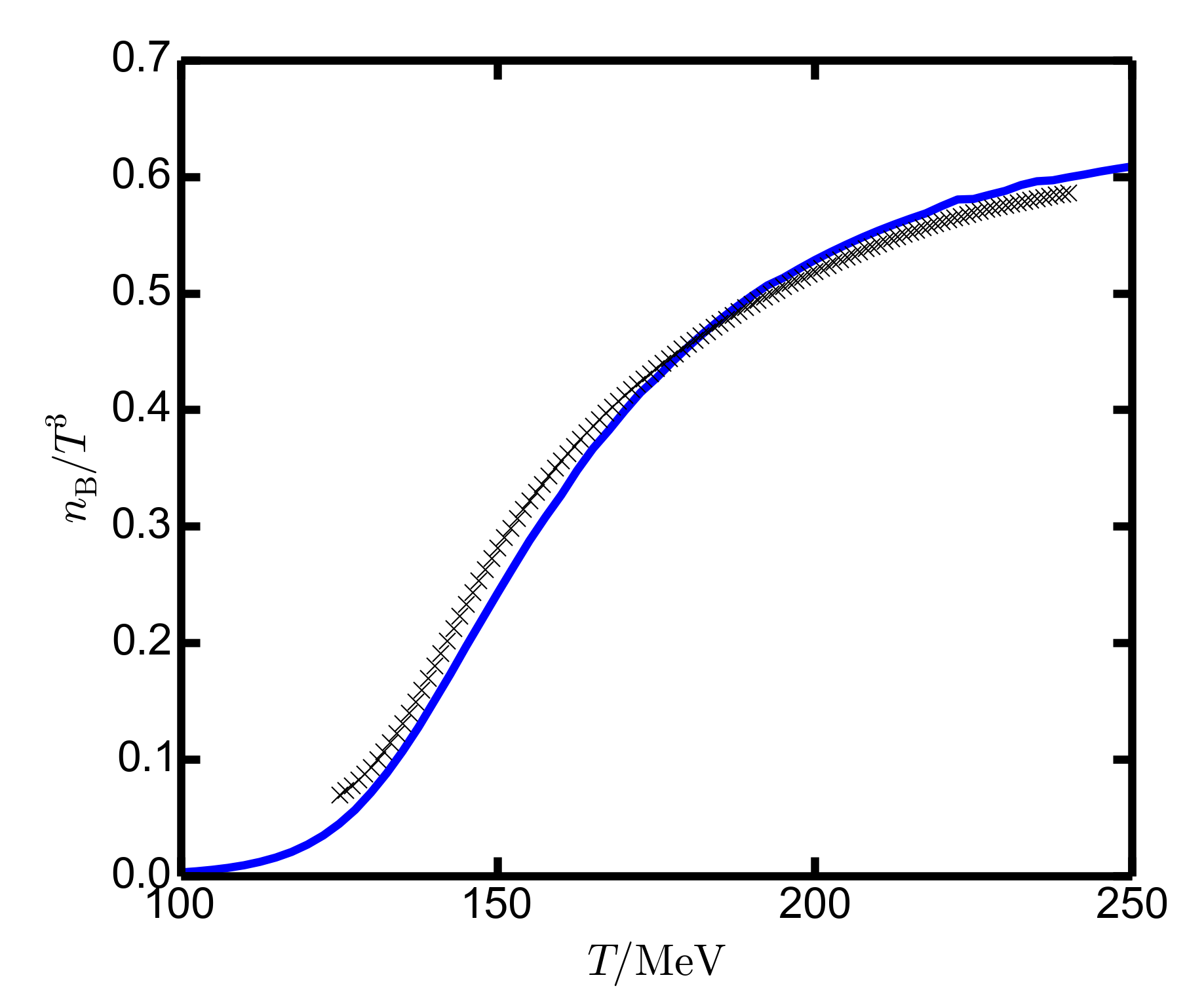}  
\includegraphics[width=0.25\columnwidth]{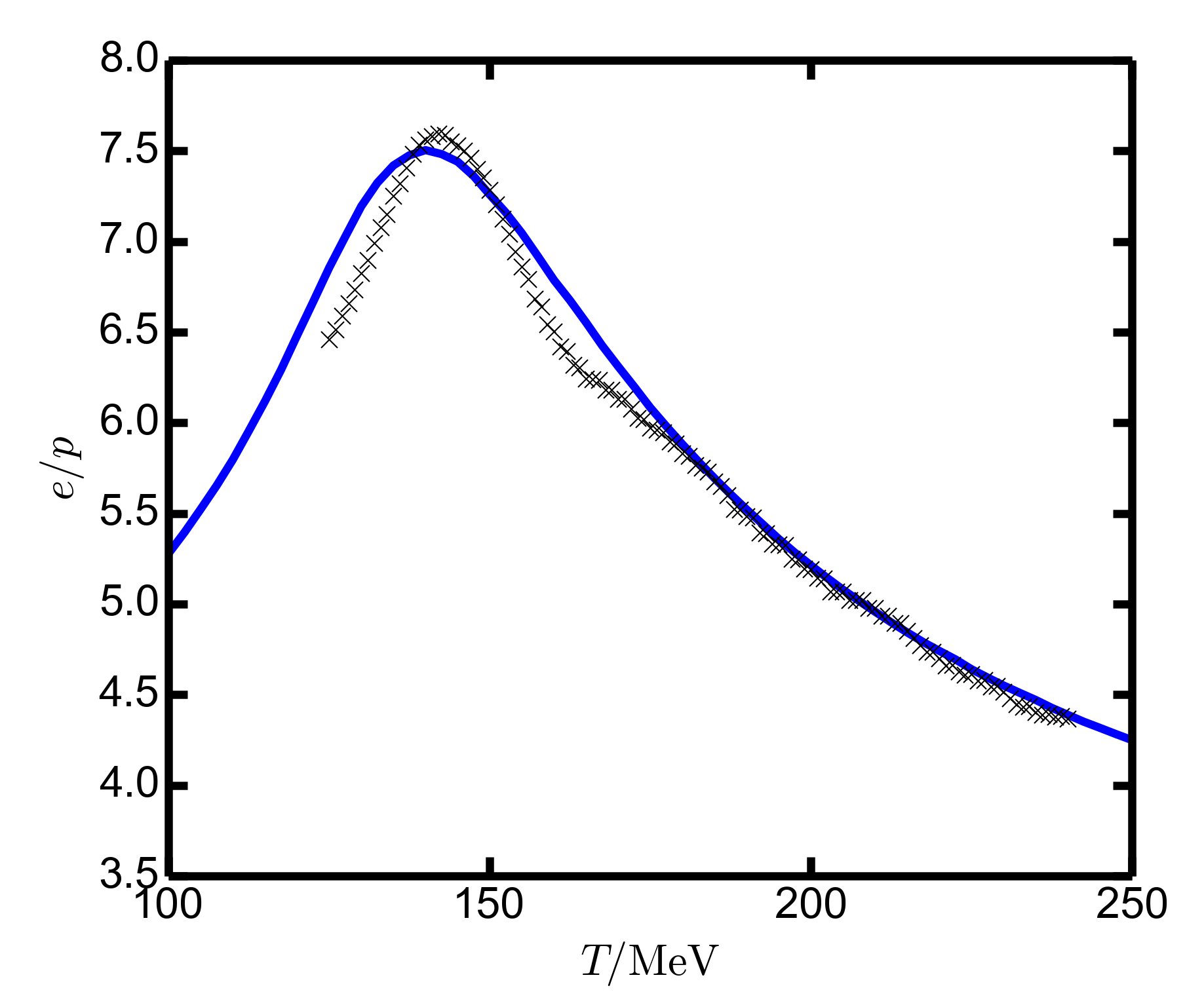} \\
\includegraphics[width=0.25\columnwidth]{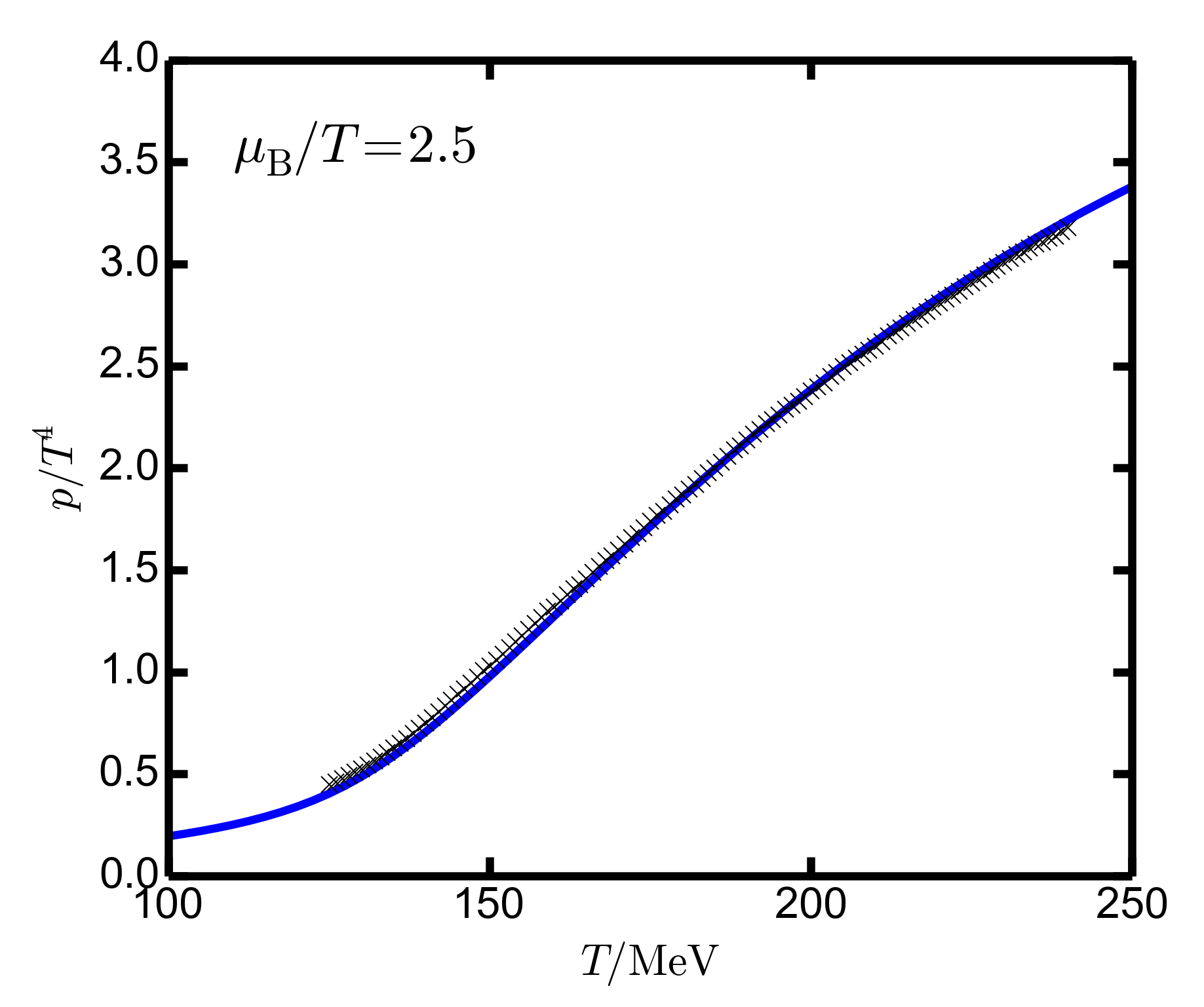}
\includegraphics[width=0.25\columnwidth]{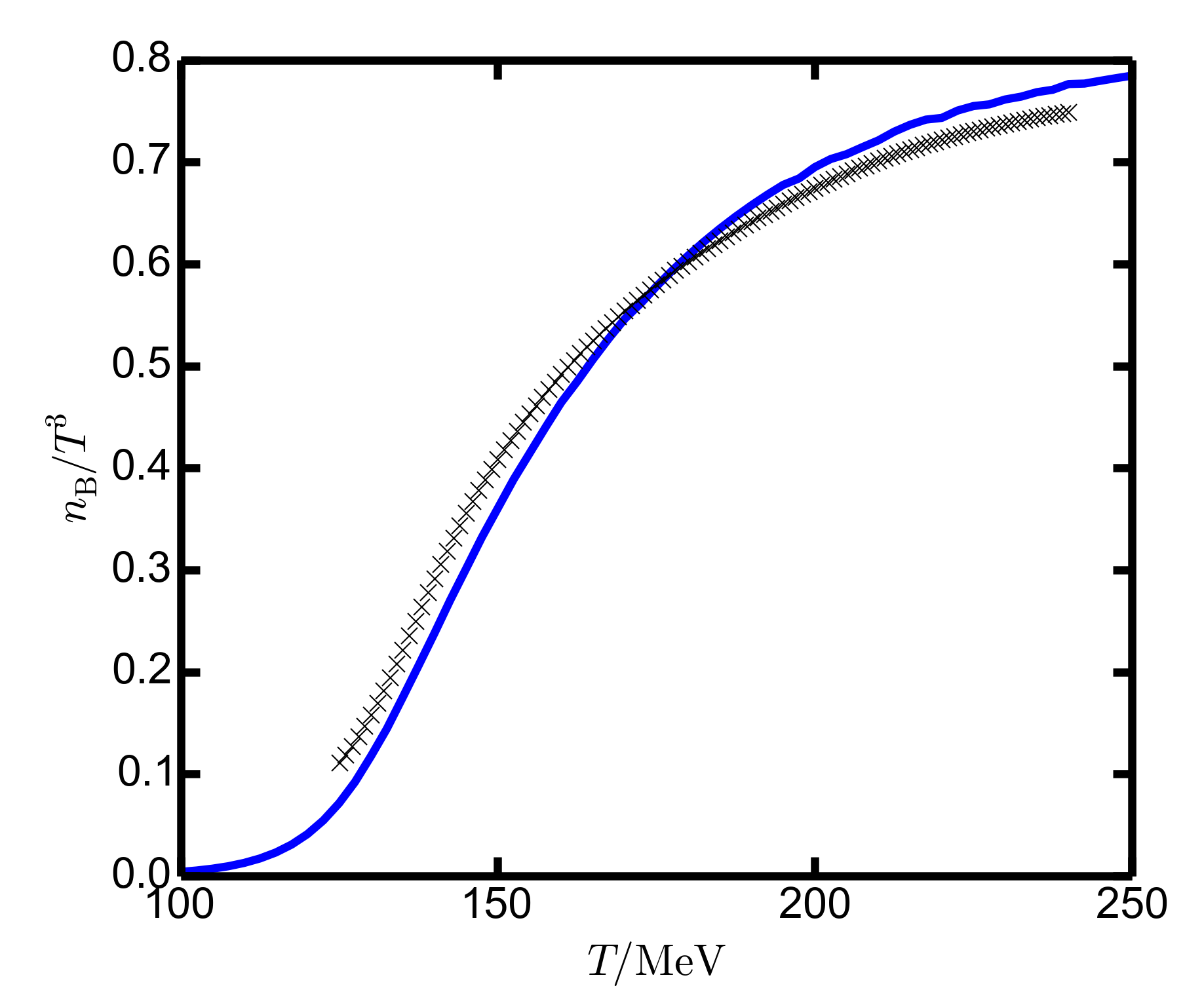}  
\includegraphics[width=0.25\columnwidth]{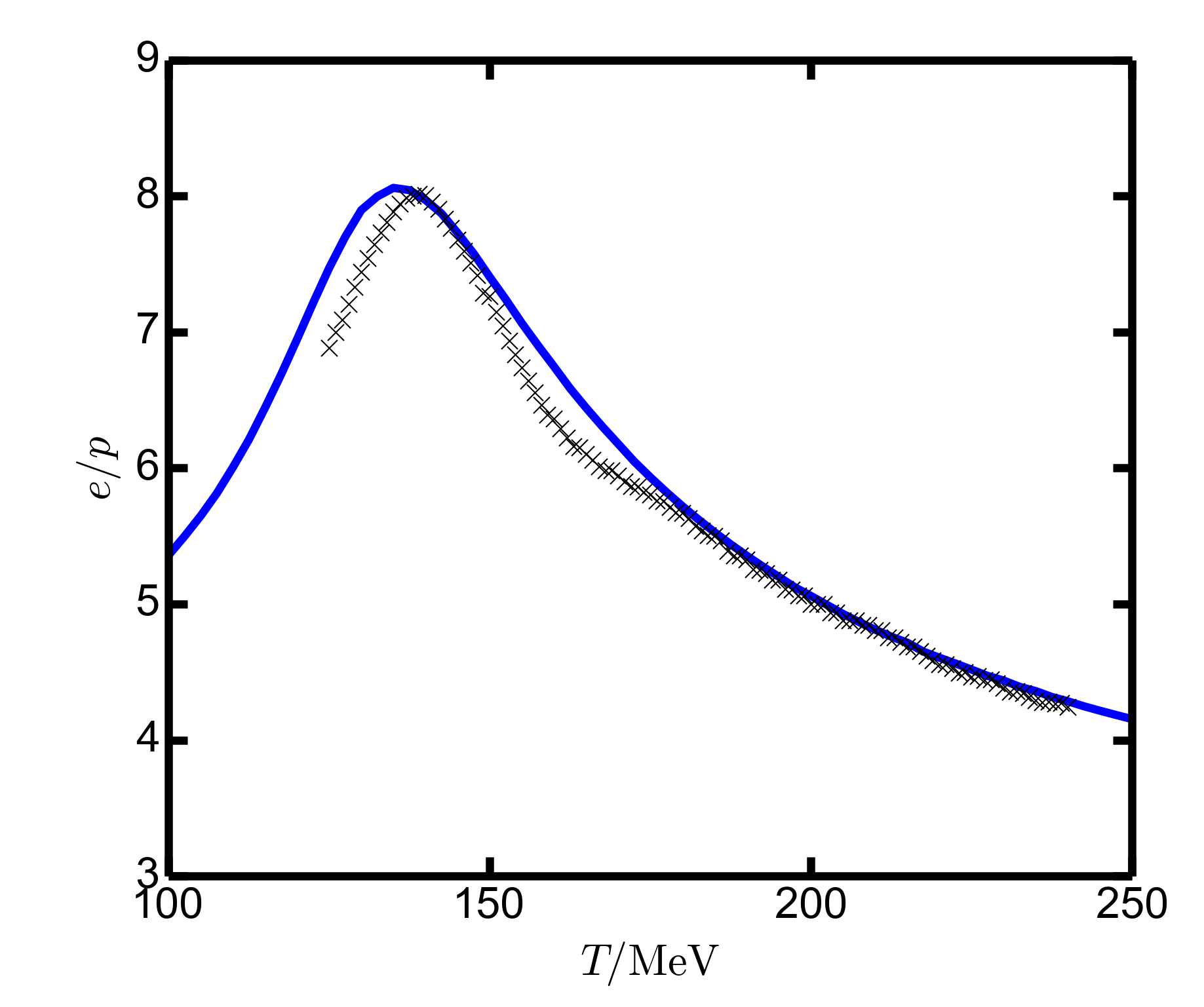}   \\
\includegraphics[width=0.25\columnwidth]{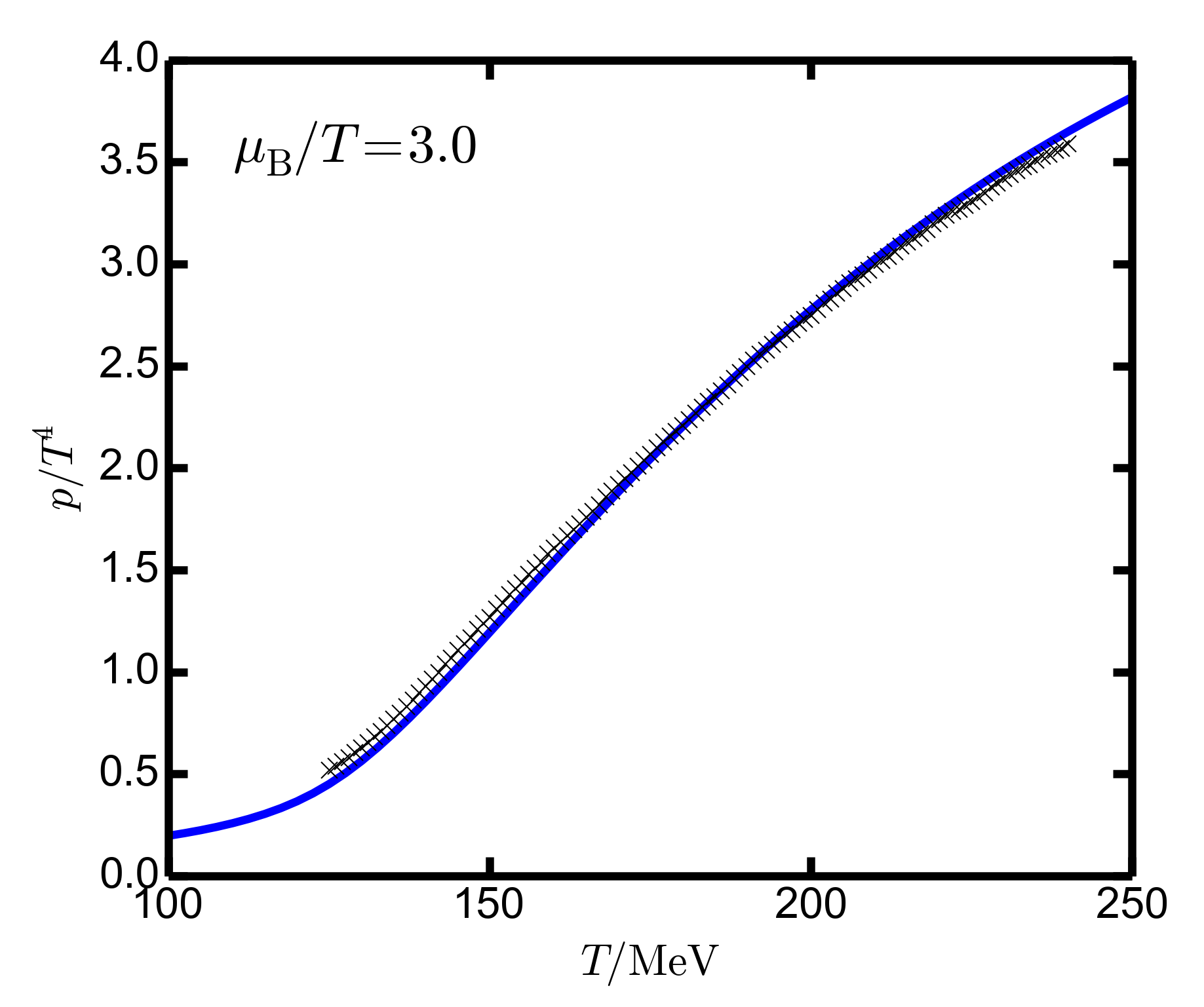}
\includegraphics[width=0.25\columnwidth]{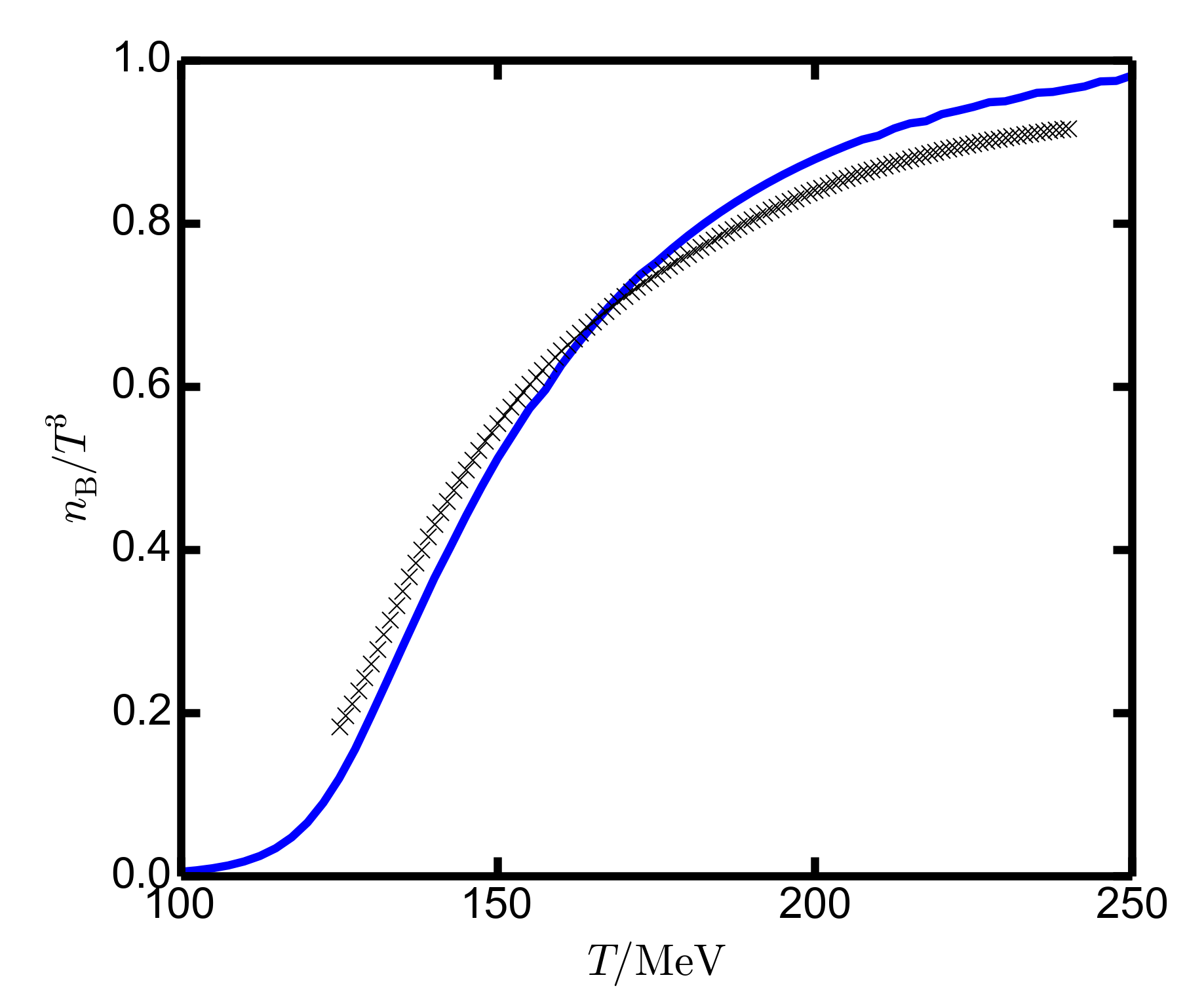}  
\includegraphics[width=0.25\columnwidth]{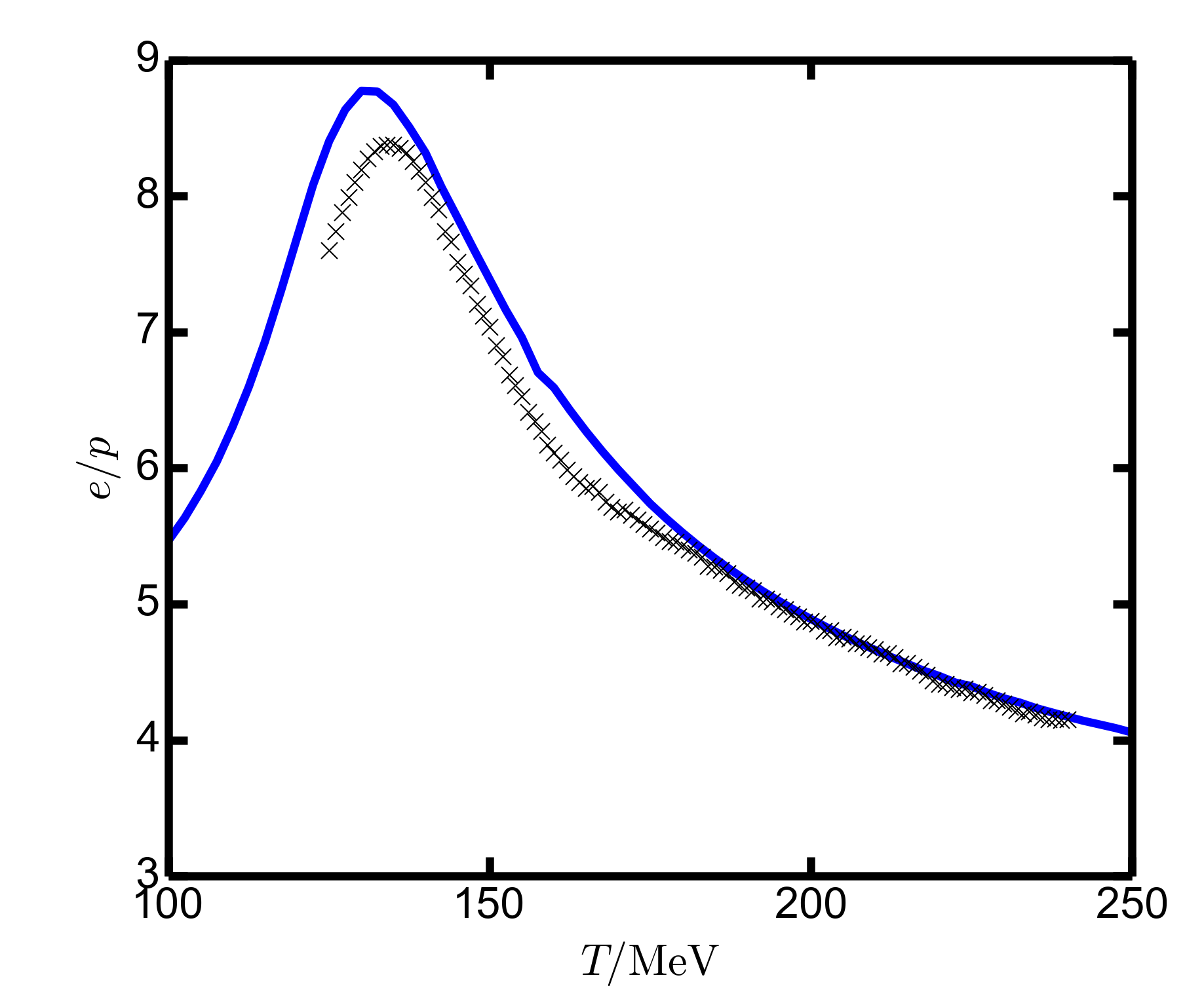} 
\vspace*{-6mm}
\caption{Comparison of the EMd model results with lattice data \cite{Borsanyi:2021sxv} (crosses)
for $\mu_B / T = n / 2$, $n = 1~\textrm{(top)}, \ldots, 6~\textrm{(bottom)}$:
$p/T^4$ (left column), $n_B/T^3$ (middle column) and $e/p$ (right column)
as a function of $T$.
Note the different scales for $n_B/T^3$ and $e/p$.
\label{fig:comparison} 
}
\end{figure}

Despite the richness of the data set \cite{Borsanyi:2021sxv}, only in scarce cases a direct comparison
with our EoS results exhibited below in various figures below is possible. 
We therefore present in Fig.~\ref{fig:comparison} a more detailed comparison focused
on $T \in [100, 250]$~MeV and for $\mu_B / T = 0.5, \ldots ,3.0$. While the scaled pressure (left column)
seems to perfectly agree, the scaled baryon density (middle column) points to some tension,
in particular for larger values of $\mu_B/T$, 
as already noticed in \cite{Grefa:2021qvt} and stressed in \cite{Cai:2022omk}.
This calls for an improved dynamical coupling ansatz.

However, we could also argue in a failure 
of the model when extending it into the confinement region,
in particular towards lower temperatures being of relevance for compact star physics.
This is in line with statements in \cite{Grefa:2021qvt}.
Thus, one could scrutinize the implications of the critical end point and related first-order phase transition
which emerge in the EMd model for parameterizations employed hitherto in the literature.
Reference \cite{Grefa:2022sav} sharpens the above warning,\footnote{
``$\ldots$, the fact that the present EMD model is in good
quantitative agreement with the latest lattice QCD data
at finite baryon density does not automatically guarantee
that the predictions made for regions of the QCD phase diagram well beyond the reach of current lattice simulations
are phenomenologically reliable. 
Indeed, the fact that the EMD model of Ref.~\cite{Cai:2022omk}
is also able to obtain a good quantitative agreement with lattice QCD thermodynamics at
zero and finite baryon density, while still predicting the
QCD CEP at a significantly different location than in our
model, shows that the available lattice data is not enough
to strongly constraint such a prediction in the EMD class of holographic models." 
}
further enhanced by statements in \cite{Jokela:2024xgz}.

Completing the discussion of Fig.~\ref{fig:comparison}, we mention that
$s/T^3$ (not displayed) obeys a similar good agreement with data as $p/T^4$.
The resulting energy density, in particular, when displayed as ratio $e/p$ (right column), again evidences some
deviations: the EMd model does not reproduce the apparent structures of the data when ignoring the error bars;
including them (not displayed) make such structures much less pronounced.
Note the slight increase of the peak of $e/p$ and its left-shift with increasing values of $\mu_B / T$.
Recap also the relations to the conformality measure $e/p = (\frac13 - \Delta)^{-1}$
and the scaled trace anomaly $(e - 3 p)/T^4 = (p/T^4) (e/p -3)$.

The dimensionless susceptibility $\chi_2 (T)$ is compared successfully to precision lattice data in figure 7-right
in \cite{Zollner:2023myk}.

\clearpage

\section{Numerical results: E\lowercase{o}S}\label{sect:EoS}

\subsection{CEP location and FOPT}

\begin{figure}[t!]
\includegraphics[width=0.44\columnwidth]{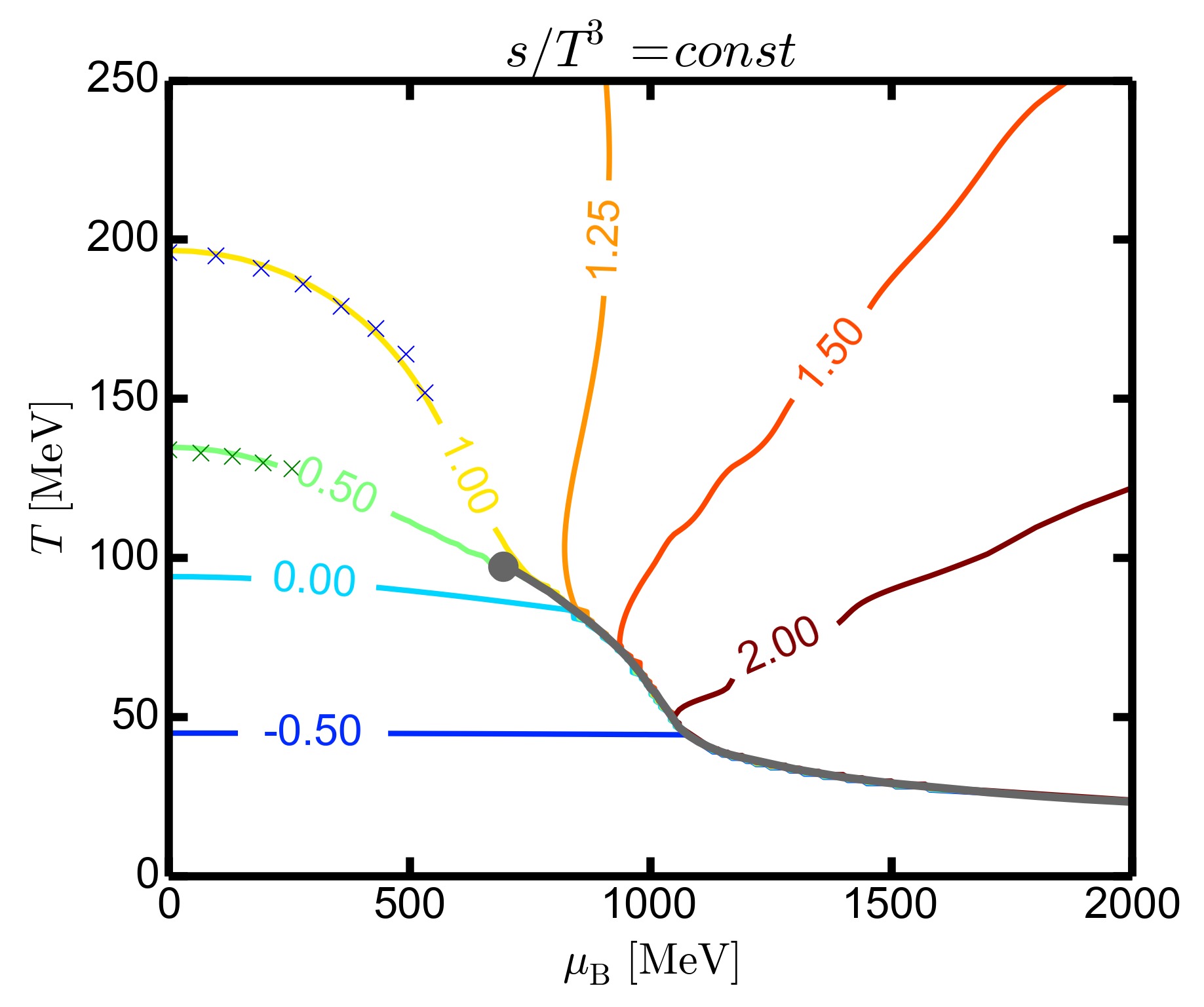}
\includegraphics[width=0.44\columnwidth]{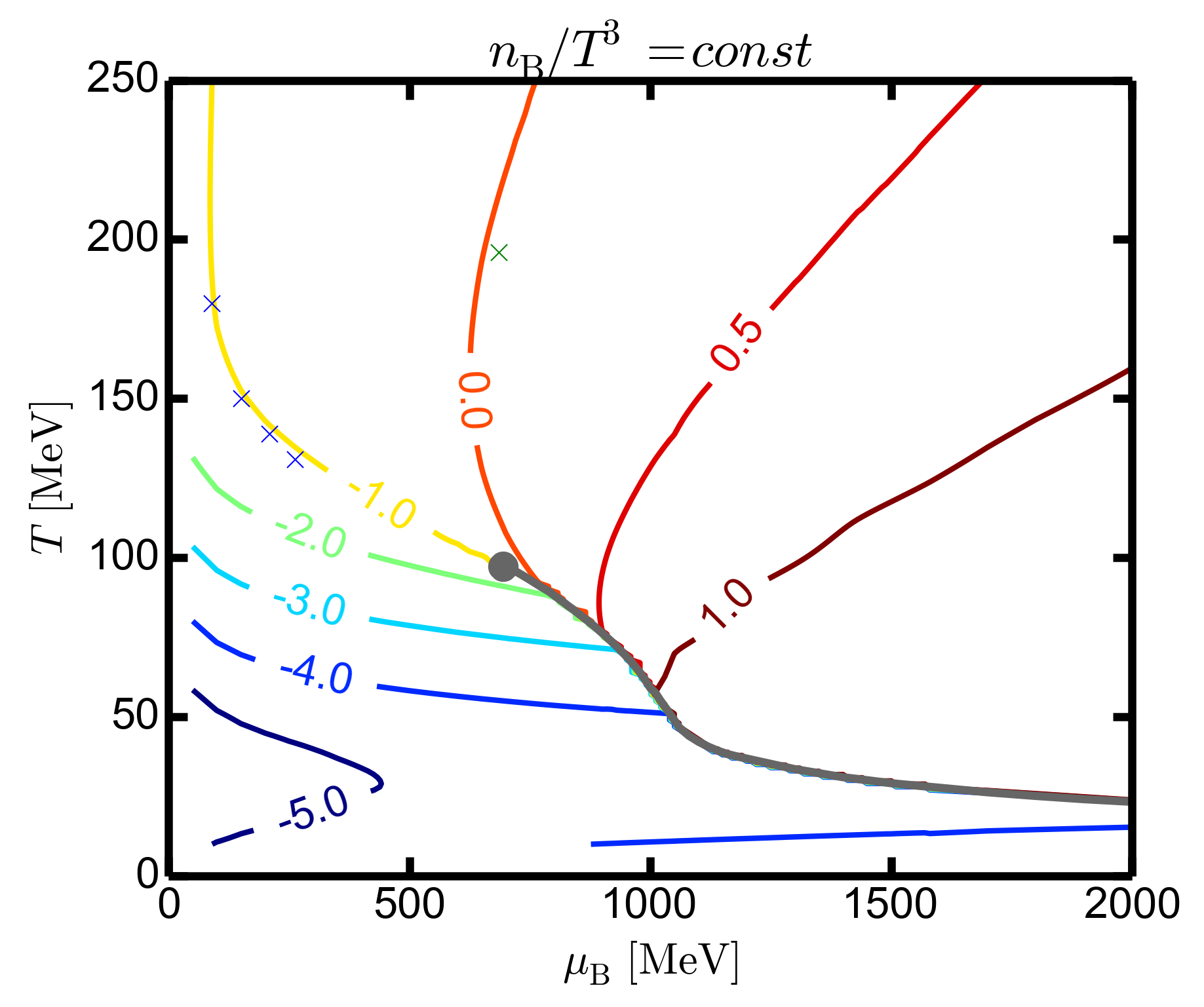} \\
\includegraphics[width=0.44\columnwidth]{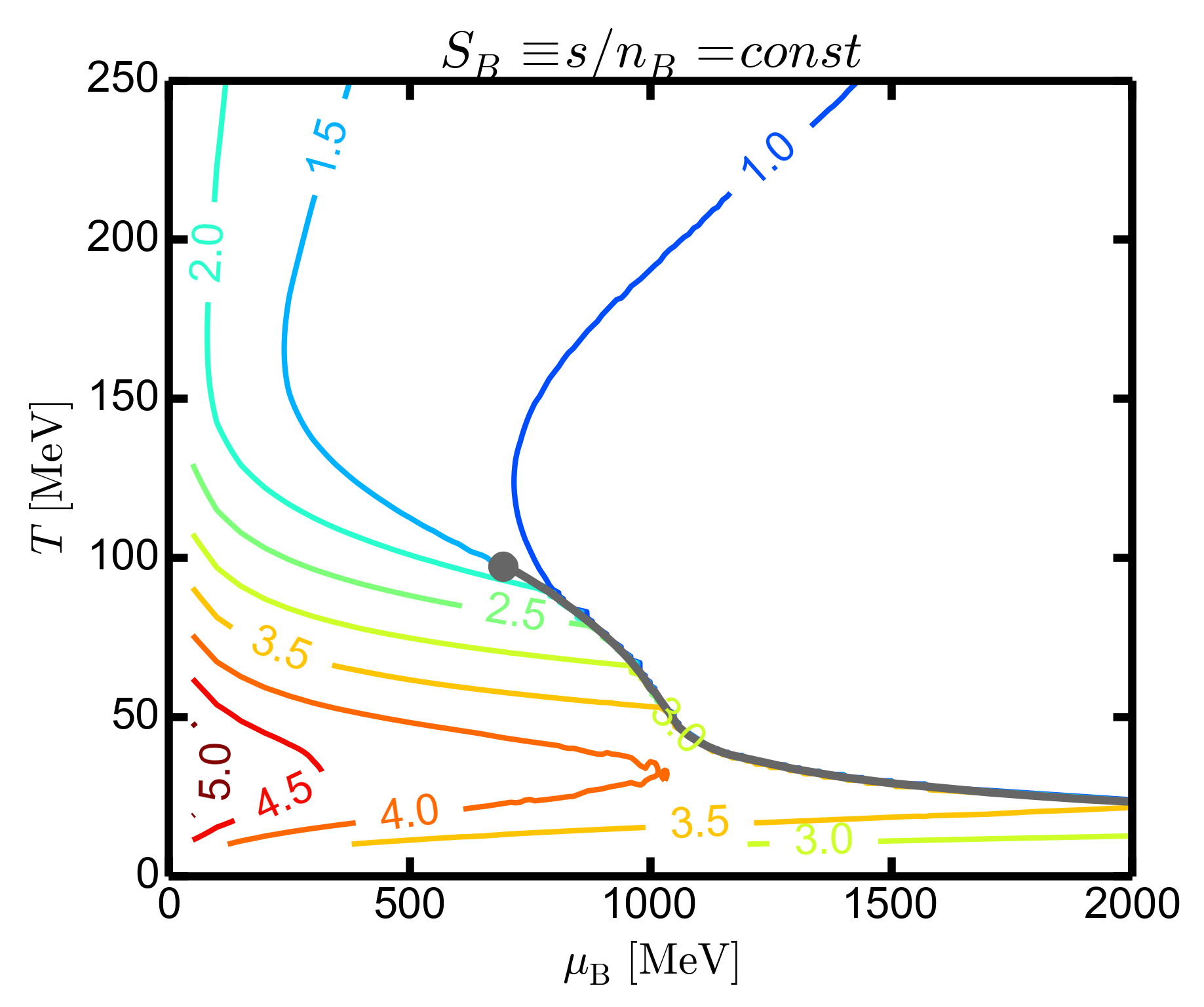}
\includegraphics[width=0.44\columnwidth]{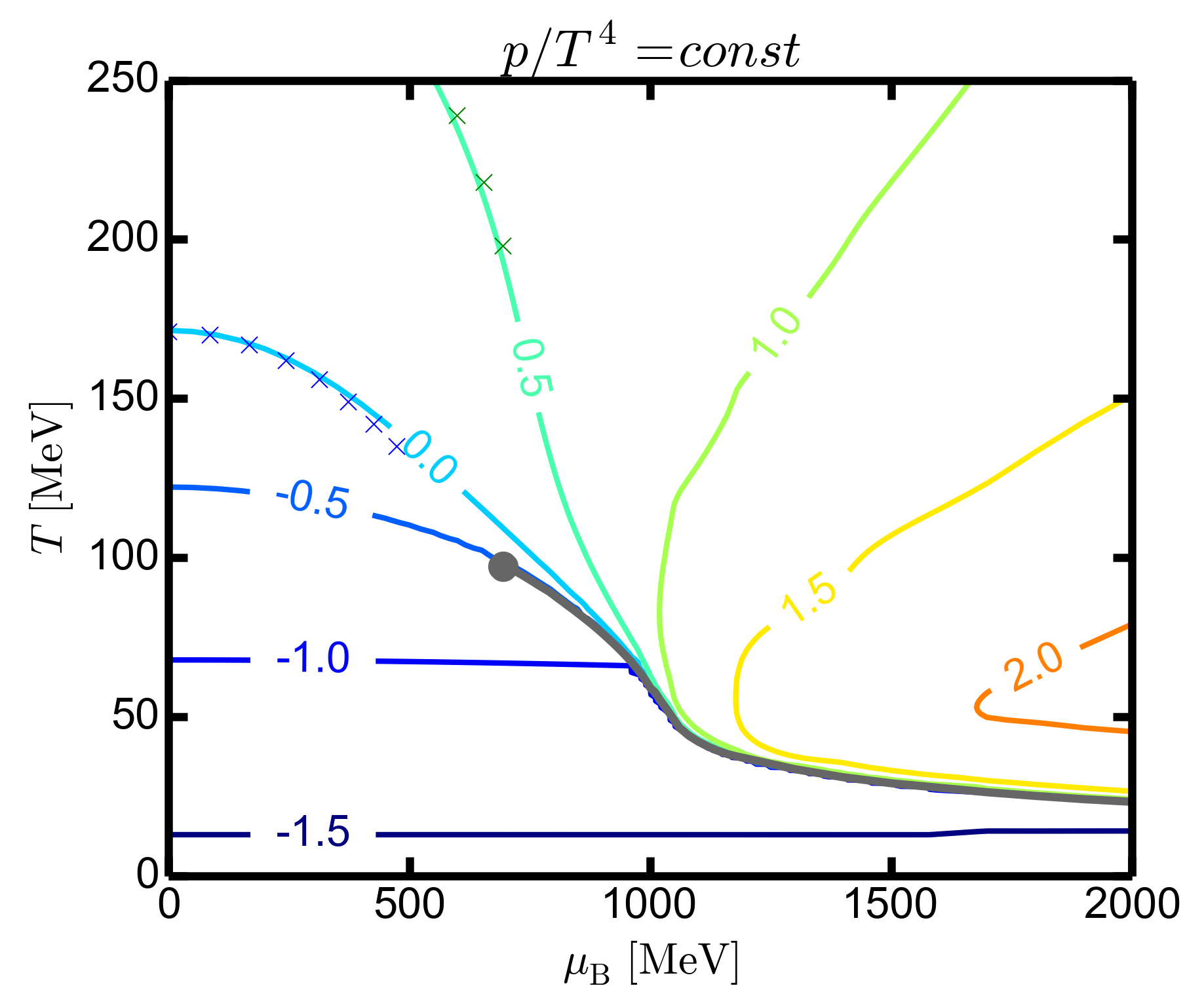}
\caption{Contour plots of scaled entropy density $s/T^3$ (left top panel), baryon density $n_B/T^3$ (right top panel) 
entropy per baryon $s/n_B$ (left bottom panel, relevant for adiabatic expansion) and pressure $p/T^4$ (right panel)
over the $T$-$\mu_B$ plane. The CEP is depicted as bullet 
and the solid black curve is for the emerging FOPT. The labeling numbers ``$N$" mean
$10^N$ of the respective quantity.
Note the weak dependence of $s/T^3$ and $p/T^4$ on $\mu_B$ left to the FOPT at $T < 100$~MeV.
The crosses depict results of the lattice QCD calculations \cite{Borsanyi:2021sxv}.
The scaled energy density, $e/T^4 = - p/T^4  + s/T^3 + (\mu_B/T) n_B/T^3$,
can be inferred from the displayed information.
\label{fig:contours_s_nB_p} 
}
\end{figure}

\begin{figure}[t!]
\includegraphics[width=0.44\columnwidth]{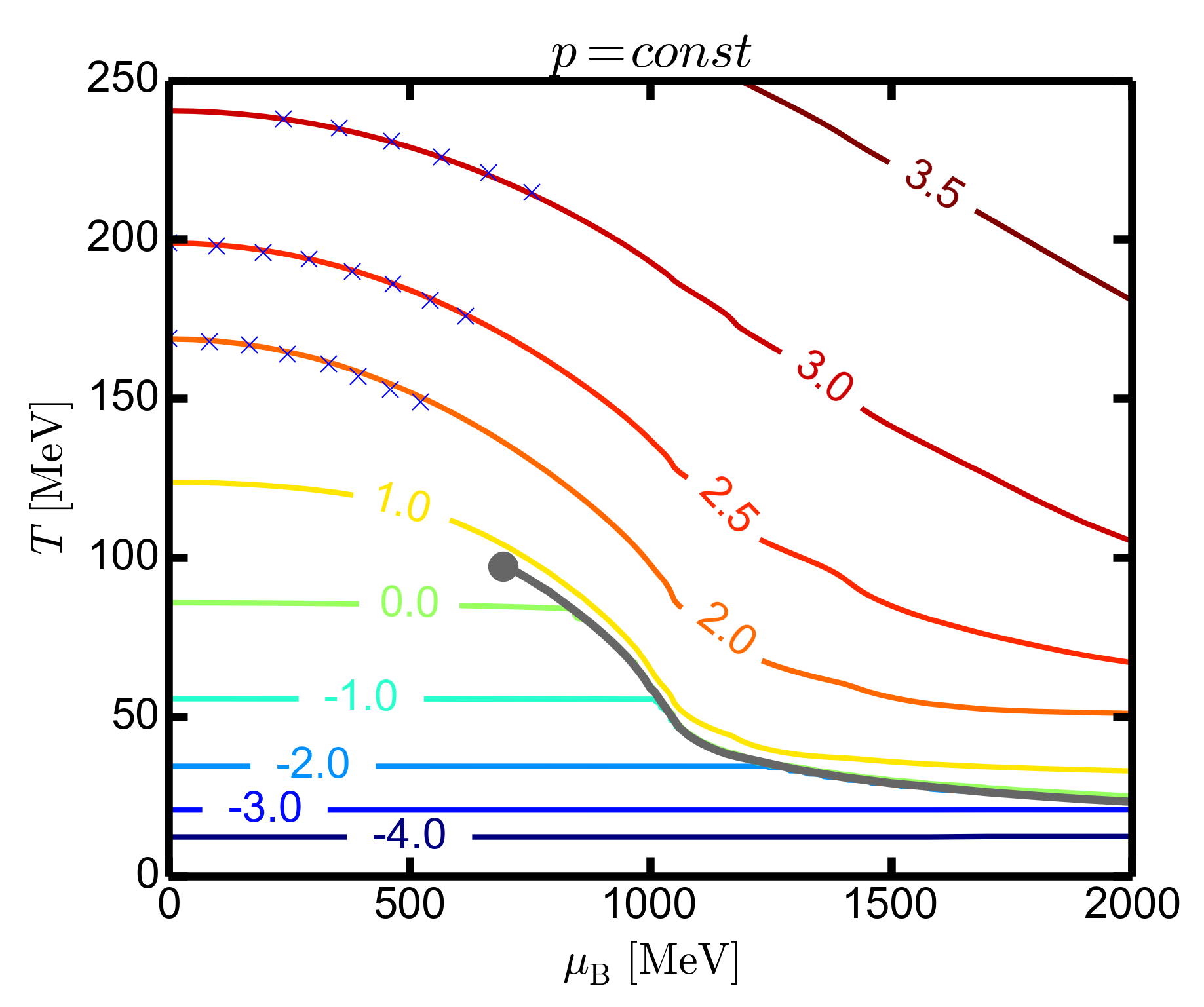}
\includegraphics[width=0.44\columnwidth]{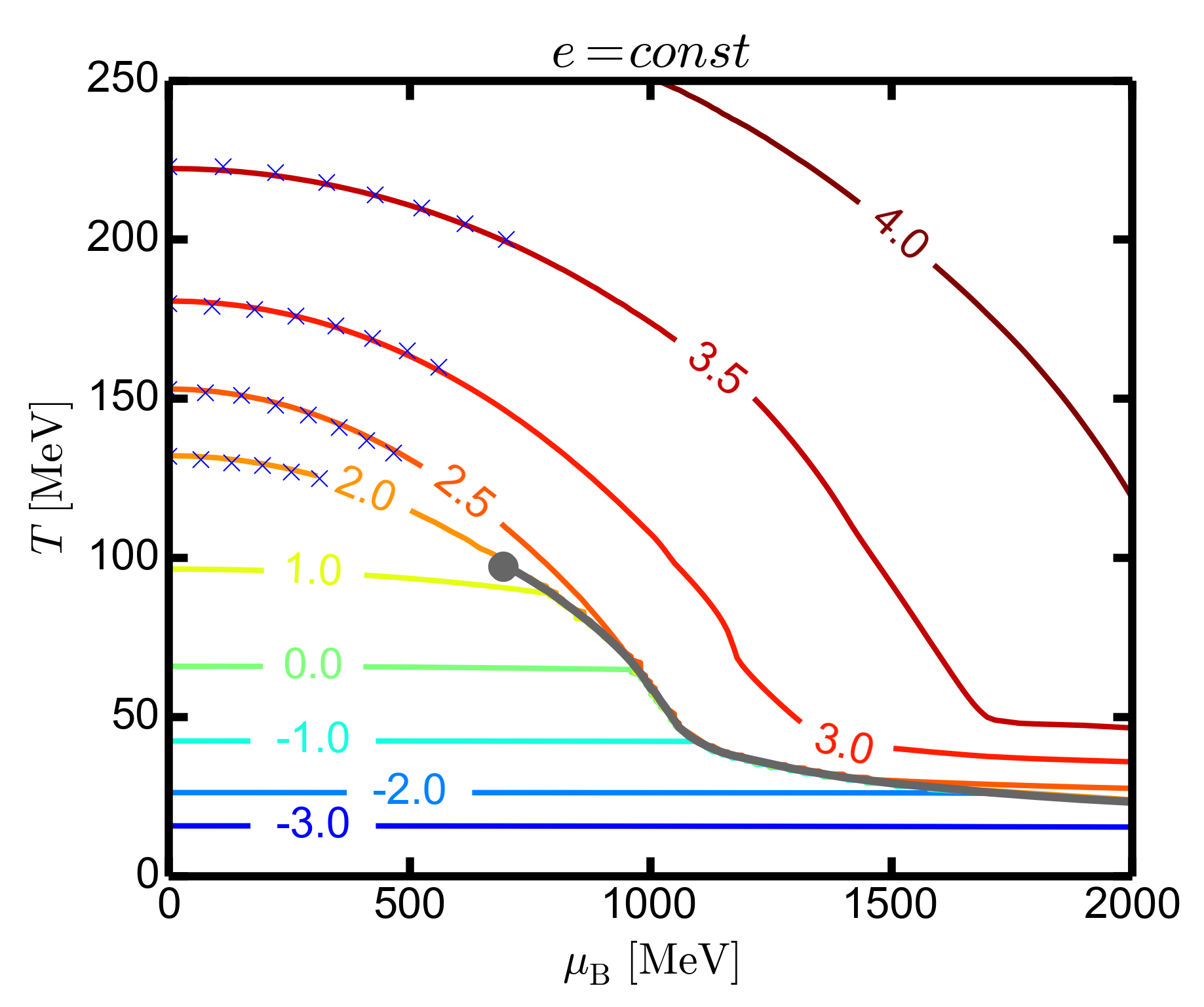}
\caption{Contour plot of the EoS as
isobars $p (T, \mu_B) = const$ (left panel)
and iso-energy density curves $e (T, \mu_B) = const$ (right panel) 
over the $T$-$\mu_B$ plane.
CEP, FOPT, line style and meaning of labeling (here in units of MeV/fm${}^3$) are as in Fig.~\ref{fig:contours_s_nB_p}. 
Note again the weak dependence on $\mu_B$ left to the FOPT at $T < 100$~MeV.
The crosses depict results of the lattice QCD calculations \cite{Borsanyi:2021sxv}.
\label{fig:contours_p} 
}
\end{figure}

The employed parameterizations of  $V$ and ${\cal G}$ facilitate a CEP with coordinates
$T_\textrm{CEP} = 97.2$~MeV and $\mu_{B \, \textrm{CEP}} = 694.7$~MeV 
(fat bullet\footnote{Similar CEP locations are obtained in
\cite{DeWolfe:2010he,Grefa:2021qvt,Critelli:2017oub,Knaute:2017opk}. 
Interestingly, QCD-functional methods
deliver fairly consistent values, cf.\ \cite{Gao:2020fbl},
and for finite-volume effects \cite{Bernhardt:2021iql}.
CEP and FOPT loci are inherent in the present model, steered by the parameterizations of
$V$ and $\mathcal{G}$, in contrast, e.g.\ to the approach in \cite{Karthein:2021nxe,Parotto:2018pwx}
which allows for a free choice.})
and a FOPT curve $T_\textrm{FOPT} (\mu_B)$ as exhibited in Figs.~\ref{fig:contours_s_nB_p}
and \ref{fig:contours_p}.
Remarkably, the FOPT curve displays near to CEP a concave shape
(as in \cite{DeWolfe:2010he,Grefa:2021qvt,Critelli:2017oub,Knaute:2017opk};
note the agreement of $T_\mathrm{FOPT} (\mu_B = 1000~\mathrm{MeV}) \approx 55~\mathrm{MeV}$ 
with figure 11 in \cite{Grefa:2021qvt})
which however turns for larger values of $\mu_B$ into a convex shape, 
obviously asymptoting the $\mu_B$ axis with dramatic consequences for the cool EoS.
Our numerically accessible domain is by far larger than that of  \cite{Grefa:2021qvt}.  

\subsection{Scaled entropy, density, pressure and specific entropy}

The contour plots of $s/T^3$, $n_B/T^3$, $s/n_B$\footnote{
The contour plot of $n_B/T^3$ (right top panel)
can be used to construct the unstable region in a $T$ vs.\ $n_B$ diagram:
A given point $(T, \mu_B)$ on the FOPT curve $T_\mathrm{FOPT} (\mu_B)$ leads to two values of the
density, $n_B^\pm (T) := n_B (T, \mu_B )\vert_{T = T_\mathrm{FOPT} (\mu_B) \pm \epsilon}$.
The unstable region has a maximum at $T_\mathrm{CEP}$, where $n_B^\pm (T_\mathrm{CEP})$ merge.
The l.h.s.\ flank, i.e.\ for the smaller values of $n_B (T)$, may be more or less steep, depending
on details of the underlying model set-up.
Figure 4 in \cite{Knaute:2017opk} exhibits that the slopes of isentropes are either so to by-pass  the
unstable region or enter \underline{and} exit it for not too small values of $s/n_B$. 
As noted in \cite{Knaute:2017opk}, the examples
in \cite{DeWolfe:2010he,DeWolfe:2011ts} are such to enter also the unstable region on the
l.h.s.\ without graceful exit. Subtle deformations of the isentropic trajectories and/or the l.h.s.\ 
of the unstable region (e.g.\ by a bumpy structure) may cause a more involved picture with
by-passing and entering-only and entering-and-exiting trajectories.
}
and $p/T^4$ exhibited in Fig.~\ref{fig:contours_s_nB_p}
point to a mysteriously weak dependence of  $s/T^3$ and $p/T^4$ on $\mu_B$ left/down to the FOPT.
The quantities $s/T^3$ and $n_B/T^3$ and $s/n_B$ as well 
jump across the FOPT, while $p/T^4$ is continuous. The r.h.s.\ continuations
of the l.h.s.\ iso-lines, e.g.\ $p/T^4 = 10^{-0.5}$ and $10^{-1}$, are not visible on the displayed scale;
they are squeezed in a narrow corridor between FOPT curve and the hardly visible curve $p/T^4 = 10$.
(See Appendix \ref{sect:appC} for some details.)
The agreement of the EMd model results and lattice data \cite{Borsanyi:2021sxv} 
(crosses, error bars are ignored here and below),
whenever possible, looks near-perfect (see also figure 8 in \cite{Zollner:2023myk}).

We emphasize the pattern of isentropic trajectories, $s/n_B = const$ (left bottom panel),
which points to a non-monotonic specific entropy at the low-temperature side
in the proposed classification of \cite{Pradeep:2024cca}. The pattern is qualitatively analog to the one
of \cite{DeWolfe:2010he,DeWolfe:2011ts} but different to that found in \cite{Knaute:2017opk},
where ``outgoing" isentropic trajectories are attributed to ``incoming" trajectories across the
FOPT upon adiabatic expansion.
It seems that various ``good fits" of the same lattice data can deliver quite different
contours of $s/n_B = const$, even we did not attempt here a dedicated $\chi^2$ optimization 
with our dilaton potential and dynamical coupling parameterizations.
Curves $e(T, \mu_B) = const$ are determined by
$\frac{\mathrm{d} T(\mu_B)}{\mathrm{d} \mu_B}\vert_{e = const} = - (T \partial s/\partial \mu_B + \mu_B \partial n_B/\partial \mu_B)/
(T \partial s /\partial T + \mu_B \partial n_B/\partial  T)$.

\subsection{Isobars and iso-energy lines}

The pattern of the isobars, see left panel in Fig.~\ref{fig:contours_p}, strongly deviates from the naive expectations
shown in Fig.~\ref{fig:Fig1} for the toy model. Neither left to the FOPT nor right to the FOPT,
one sees a near-vertical dropping of the curves $p=const$. Instead, the mysterious low--temperature-low-$\mu_B$
behavior of the scaled pressure exhibited in the right panel of Fig.~\ref{fig:contours_s_nB_p} is retained, and, 
most importantly for our goals, the FOPT curves seem to repel the isobars. Thus, a smooth continuation of
curves $p=const$ to the $\mu_B$ axis is hindered in the explored range. A warm EoS at $T > 50$~MeV
is conceivable by our approach, but the envisaged cool EoS at small or zero temperature remains elusive.  
Also in the current case, the agreement with lattice data \cite{Borsanyi:2021sxv} (crosses) looks fine.  

\subsection{Warm EoS}

\begin{figure}[t!]
\includegraphics[width=0.44\columnwidth]{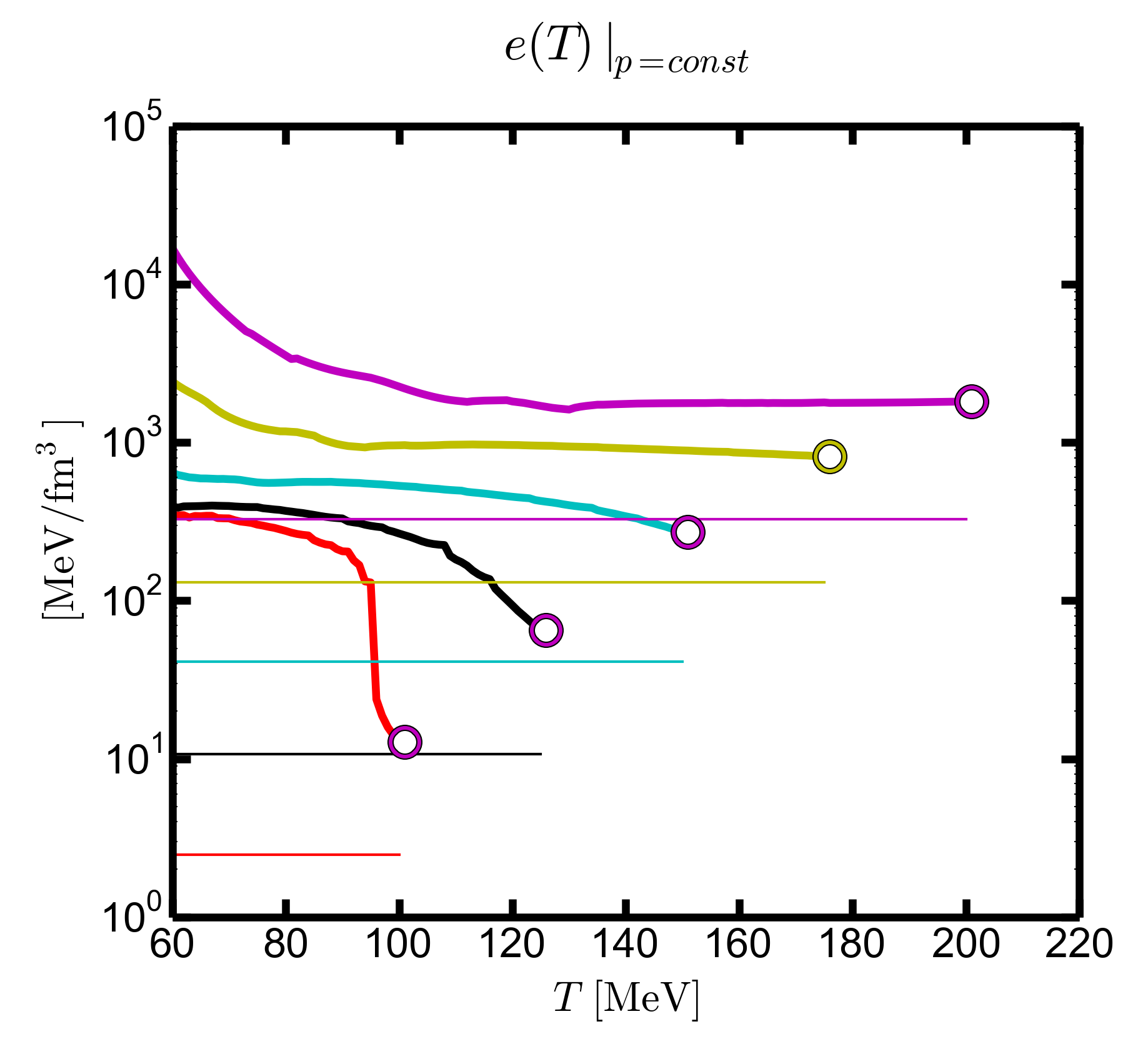}
\includegraphics[width=0.44\columnwidth]{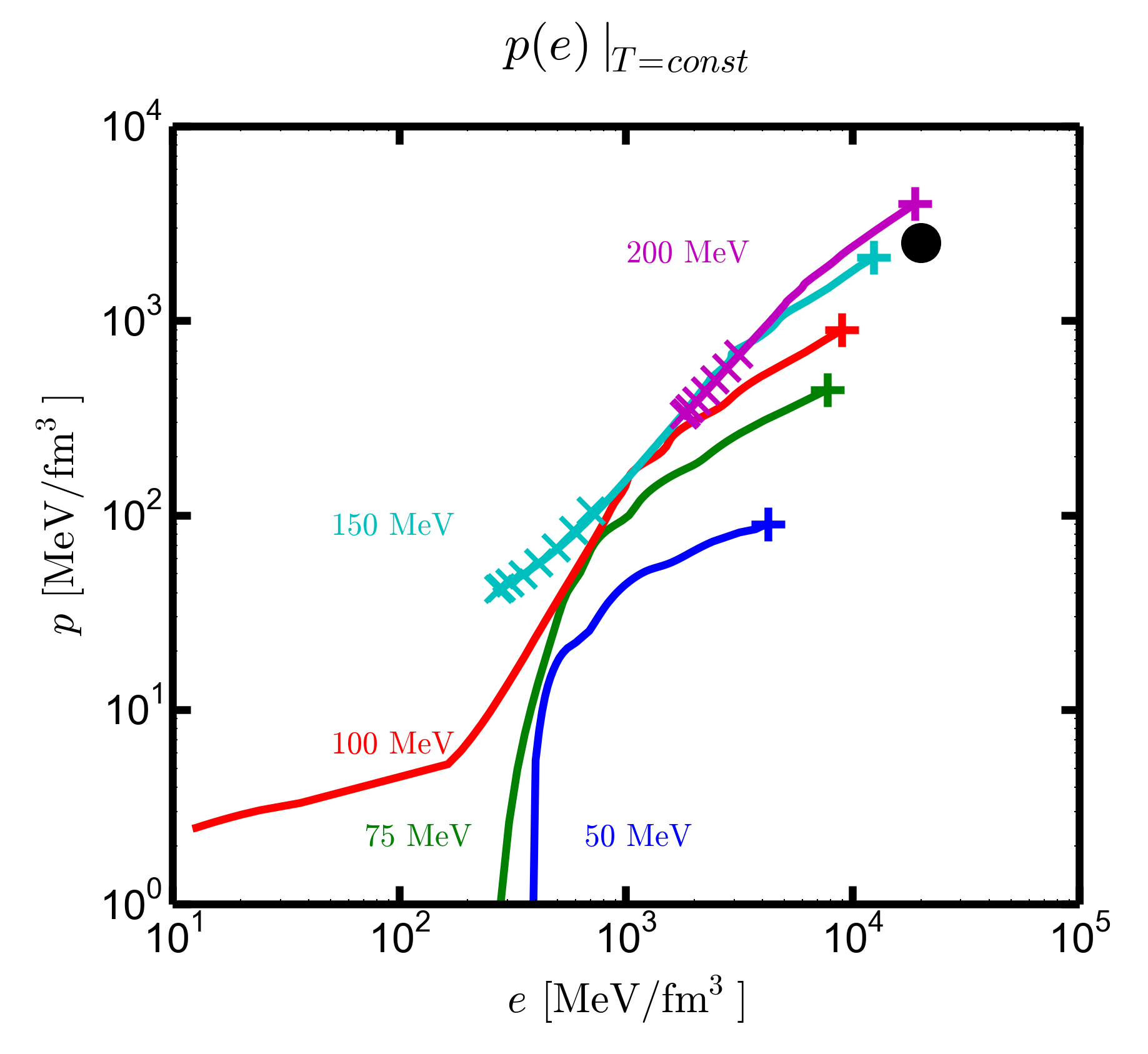}
\caption{Left panel: 
Energy density $e$ (solid curves) as a function of temperature $T$ along the ``save" isobars $T(\mu_B)\vert_{p=p_0}$ 
(see Fig.~\ref{fig:contours_p}-left)
for various values of $T_0 =  125$ (black), 150 (cyan), 175 (yellow), and 200~MeV (magenta)
and thus $p_0 = p(T_0, \mu_B = 0)$. In addition, the case of a ``less reliable" isobar with
$T_0 = 100$~MeV is also displayed (red).
The r.h.s.\ endpoints ("o") are for $\mu_B = 0$, 
both for $e$ and 
pressure $p = p_0$ (horizontal  thin lines, same color code as the corresponding energy density).
The difference of $e$ and $p$ (both in units of MeV/fm${}^3$)
in the employed log scale delivers directly $e/p$ as a function of $T$
along the respective isobar. Equally well, $e$ and $p$ for a selected constant value of $T$ can be read off,
thus providing the iso-thermal EoS $p(e)\vert_{T = const}$, exhibited in the \underline{right panel} for
various temperatures as provided by labels. 
The r.h.s.\ endpoints "+" are for $\mu_B = 2000$~MeV.
One could also combine the results of Fig.~\ref{fig:contours_s_nB_p} along cuts of $T = const$
to arrive at the same picture.
The crosses depict results of the lattice QCD calculations \cite{Borsanyi:2021sxv} in both panels.
The bullet depicts the onset point of the perturbative QCD regime  for $T = 0$.
Nuclear many-body theory is expected to apply below left bottom corner. 
\label{fig:EoS} 
}
\end{figure}

To illustrate further features of the obtained EoS we exhibit in Fig.~\ref{fig:EoS} (left panel)
energy density $e$ and pressure $p_0$ as a function of temperature $T$ along the isobars $T(\mu_B)\vert_{p=p_0}$ 
(or $\mu_B (T)\vert_{p = p_0}$ by inversion)
for various values of $T_0$ which generate $p_0 = p(T_0, \mu_B = 0)$.
That is the EoS in parametric form, $e(T, \mu_B(T)\vert_{p = p_0})$ and $p(T, \mu_B(T)\vert_{p = p_0})$,
i.e.\ from this information one can infer directly $e/p$ due to the log scale  (note the relation to the
trace anomaly measure \cite{Fujimoto:2022ohj}
via $e/p = 1/(\frac13 - \Delta)$) or $p(e)$ (see right panel).
As mentioned above, the envisaged access to the small-$T$ region is hampered by the failed approach of isobars
down to the baryon-chemical potential axis, even for the ``safer" isobars emerging from or running through the 
$T$-$\mu_B$ region uncovered by the lattice QCD results \cite{Borsanyi:2021sxv}.   
We stress the enormous stiffness of the EOS on the r.h.s.\ of the FOPT.
In line with the peculiar features of the EoS at $T < T_\mathrm{CEP}$ and matter properties
at l.h.s.\ of the FOPT is the huge energy density jump.

The connection of the holographic EoS with the nuclear physics-based EoS at acertain matching point 
requires a special treatment, beyond the goal of the present paper. Analogously, the transit to the perturbative QCD
regime above about $4 \times 10^3$~MeV/fm${}^3$ (pressure) and $1.3 \times 10^4$~MeV/fm${}^3$ (energy density)
also requires separate work. As a guide line, one could follow the construction of a hybrid EoS in
the spirit of figure 15 in \cite{Jarvinen:2021jbd} and plug in the present holographic EoS as one building block
among others.
It happens however that, at given energy density, e.g.\ $10^3$~MeV/fm${}^3$, our pressure seems too low
for $T < 100$~MeV in comparison with currently advocated credible EoSs, 
cf.\ figure 1 in \cite{Ecker:2022dlg,Annala:2021gom}.

\section{Conclusions and Summary}\label{sect:conclusions}

Einstein-Maxwell-dilaton (EMd) models became quite popular during the last years. They are aimed at
getting a quantified hint on the location of a (hypothetical) critical end point (CEP) and emerging
first-order phase transition (FOPT) curve in the phase diagram of QCD. The experimental search in
relativistic heavy-ion collisions and conjectured impact on neutron star merger dynamics and
compact (neutron) star configurations up to sources of stochastic gravitational waves formed in the 
cosmological confinement epoch 
provide strong motivation to apply the class of EMd models in a regime where direct first-principle
QCD calculations are not (yet) at our disposal.

EMd models are argued to deliver reliable information in the deconfinement regime.      
Once the dilaton potential and the dynamical coupling are adjusted, a specific EMd model parameterization
delivers as primarily quantities entropy density $s$ and
baryon density $n_B$, thus allowing to construct isobars $p = const$.
These curves $T(\mu_B)\vert_{p=const}$ map the pressure profile, given by lattice QCD data on the temperature axis,
into the temperature-baryon--chemical potential plane towards the baryon--chemical potential axis.
Supplemented by the available information on $s$ and $n_B$ to gain the accompanying energy density $e$,
the warm equation of state $p(e)$ becomes accessible. 

The model facilitates a critical end point and a peculiar first-order phase transition curve, which levels off partially
due to an imposed optionally novel side condition to suppress the appearance of an unwanted purely thermal phase transition.
Thus the pressure interval directly accessible by lattice QCD on the temperature axis is transformed to a pressure profile
at smaller temperatures: $p(T, \mu_B = 0) \mapsto p(\mathrm{smaller~}T \approx 0, \mu_B > 0)$.
We emphasize that the lattice QCD pressures in the temperature interval $T \in [125, 240]$~MeV continue (numerically)
the equation of state of strong-interaction matter towards a region relevant for compact (neutron) stars 
and their merging dynamics but not accessible
reliably by nuclear many-body theory. We also emphasize that pressures 
$p(T \notin [125, 240]~\mathrm{MeV}, \mu_B = 0)$
are largely unconstrained (and could be merely hampered by the
employed ansatz of dilaton potential and dynamical coupling of the EMd model), and therefore their mapping
$p(T, \mu_B = 0) \mapsto p(T \approx 0, \mu_B)$ is hazardous, in contrast to the valuable control of pressures
$p(T \in [125,  240]~\textrm{MeV}, \mu_B =0) \mapsto p(T \in [125,  240]~\textrm{MeV}, \mu_B >0)$ 
by lattice data with $0 < \mu_B / T \le 3.5$.

Since the EMd model with our presently employed parameterization points to a peculiar shape of the
first-order phase transition curve, one should test other ans\"atze to elucidate whether suitable side
conditions have strong impact. In a larger context, a systematic approach to reliable parameterizations of the
dilaton potential and dynamical coupling within EMd models is desirable.  
The low--temperature-low--density behavior of the model appears somewhat mysterious and could point
to the need of including explicitly fermionic degrees of freedom,
i.e.\ nucleons and their hard-core repulsion.
First-principle input would be highly welcome to constrain better the model and provide confidence at
larger densities.
For practical applications, however,
that region is uncovered by nuclear many-body approaches, and the present model should be constrained
to the high-density continuation relevant to compact stars.
The matching to perturbative QCD is a further constraint to the holographic EoS to be considered
in follow-up work towards a hybrid EoS, where eventually also local charge neutrality and 
$\beta$ equilibrium should be imposed.

\begin{appendix}

\section{A toy model of isobars}\label{sect:appD}

\begin{figure}[t!]
\includegraphics[width=0.51\columnwidth]{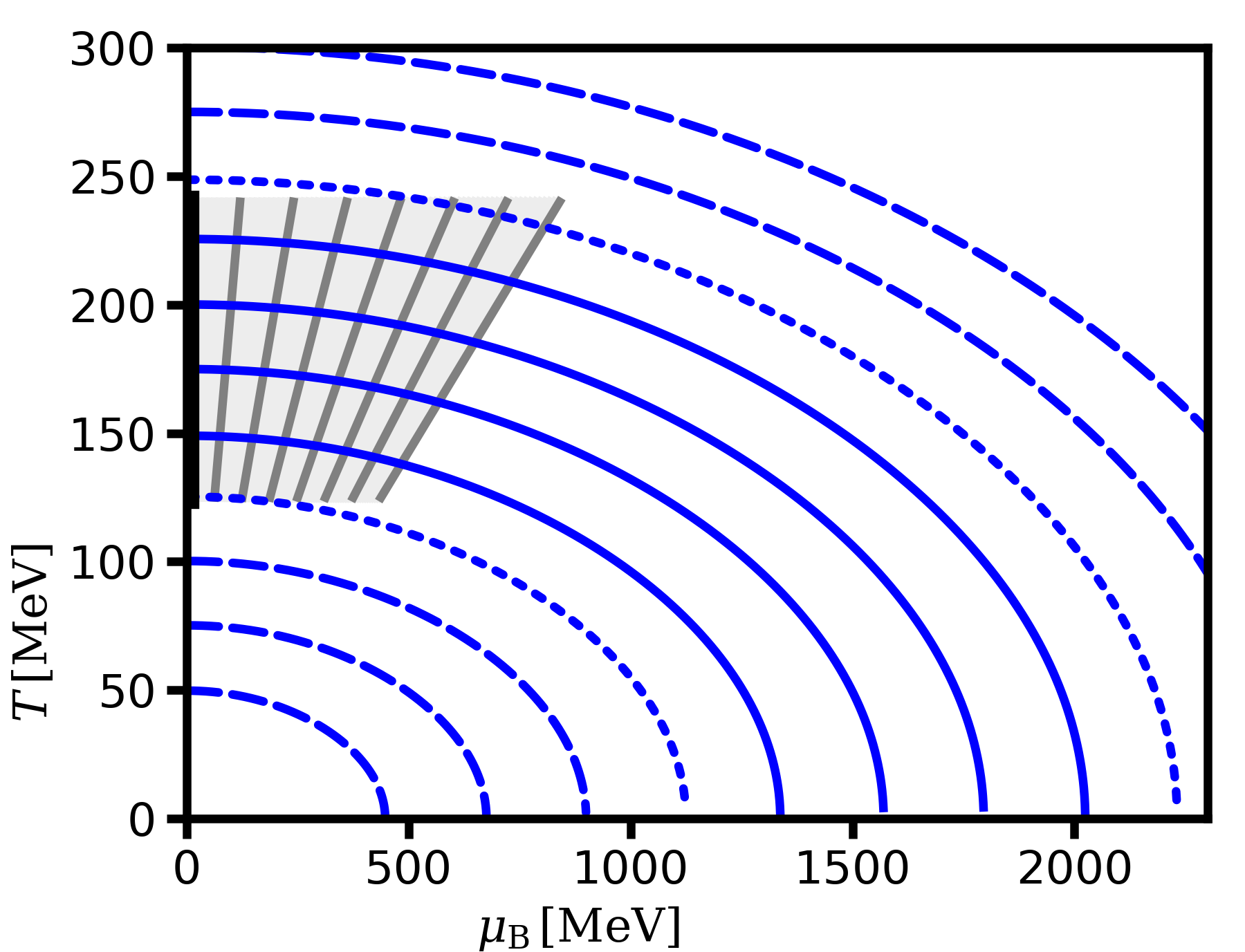}
\put(-246,88){\rotatebox{90}{ $\underbrace{\textrm{QCD input}}_{ }$ } }
\put(-99,9){ $\underbrace{ \hspace*{2.7cm} }_{\textrm{model output: $p(\mu_B)$ } }$ }
\caption{Illustration of expected isobars $p (T, \mu_B) = const$ over the $T$-$\mu_B$ plane
in a toy model.
The heavy solid bar on the $T$ axis indicates the region, where reliable QCD input data (e.g.\
$p(T)$ and $\chi_2(T)$) are at our disposal. The continuation to $\mu_B > 0$ is controlled by lattice
data in the hatched region (with sections of rays $\mu_B /T = const$ highlighted). 
Isobars not emerging from the heavy solid vertical bar or not running
a noticeable section through the hatched control region are to be considered as less reliable (dashed or dotted curves).
Irrespective of the EoS on the $T$ axis, such a mapping by ``laminar curves" $T(\mu_B)\vert_{p= const}$
(solid curves)
would allow to arrive unambiguously at the cool EoS at $T = 0$, or any other cut through the $T$-$\mu_B$ plane,
thus providing also a warm EoS for neutron star merger dynamics. 
\label{fig:Fig1}
}
\end{figure}

Figure \ref{fig:Fig1} illustrates the envisioned mapping of the hot EoS, $p(T, \mu_B=0)$,
into the $T$-$\mu_B$ plane and eventually to the $\mu_B$ axis to arrive at the warm EoS, $p(T, \mu_B)$,
and cool EoS, $p(T=0, \mu_B)$, respectively, by a toy model.
The mapping is accomplished generically by isobars, i.e.\ curves $p(T, \mu_B) = const$
determined by Eq.~(\ref{eq:T(muB)}).  
The displayed curves in Fig.~\ref{fig:Fig1} are for a toy model with 
$p = a T^4 + b T^2 \mu_B^2 + c \mu_B^4 + d$ implying
$\hat s \equiv s/T^3 = 4 a + 2 b \hat\mu_B^2$, 
$\hat n_B \equiv n_B /T^3 = 4 c \hat \mu_B^3 + 2 b \hat \mu_B$
and numerical values $b/a = 0.02738$, $c/a = 0.000154$
which refer to a non-interacting two-flavor quark-gluon medium. 
Note that, in this special case of $\hat s$ and $\hat n_B$ depending only on $\hat \mu_B \equiv \mu_B /T$,
an isobar starting at $T = T_0$ ends at $\mu_{B \, 0} = 8.79 \, T_0$. 
Isentropic trajectories, $T \propto \mu_B$ from $s/n_B = const$,
miss the typical back-bending at $T = \mathcal{O} (150~\textrm{MeV})$ 
(see figures 4  in \cite{Flor:2020fdw},
10 in \cite{Karthein:2021nxe},
11 in \cite{Parotto:2018pwx}
and 4 in \cite{Gunther:2017sxn}),
thus, evidencing the limited applicability of the toy model in the confinement region. 
In particular, a CEP and related FOPT curve are completely missed in this toy model.

\section{Details of the EMd model}\label{sect:appA}

The action (\ref{eq:EMd_action}) with the metric (\ref{eq:metric}) leads to the field equations
\begin{align}
A'' &= -\frac{1}{6}\phi^{\prime \, 2} , \label{eq:A}\\
f'' &= e^{-2A} \, \mathcal{G} \, \Phi^{\prime \, 2} - 4 A^\prime f^\prime , \label{eq:f}\\
\phi'' &= \frac{1}{f}\left(\partial_{\phi} V - \frac12 e^{-2A}\, \Phi^{\prime \, 2} \, \partial_{\phi} \mathcal{G} \right)
-\left(4A^\prime+\frac{f^\prime}{f} \right)\phi^\prime , \label{eq:phi}\\
\Phi'' &= -2A^\prime \phi^\prime - \frac{\partial_{\phi} \mathcal{G}}{ \mathcal{G}} \, 
\phi^\prime \, \Phi^\prime , \label{eq:Phi}
\end{align}
to be solved with boundary conditions
\begin{align}
A(0)    &= 0 ,  & A'(0) &= - \frac16 \left( 2 V(\phi(0)) + \Phi_1^2 \, \mathcal{G}(\phi(0)) \right) , \tag{\ref{eq:A}'}\\
f(0)     &= 0 ,  & f^\prime(0) &= 1 , \tag{\ref{eq:f}'}\\
\phi(0) &= \phi_0 , & \phi'(0) &= 
\partial_\phi V\vert_{\phi(0)} - \frac12 \Phi_1^2 \, \partial_\phi \mathcal{G}\vert_{\phi(0)} , \tag{\ref{eq:phi}'}\\
\Phi(0) &= 0 ,  & \Phi'(0) &=  \Phi_1 , \tag{\ref{eq:Phi}'}
\end{align} 
where a prime means derivative w.r.t.\ coordinate $r$.

Entropy density and baryon density follow from (cf.\ \cite{DeWolfe:2010he})
\begin{align}
s(T, \mu_B) &= \frac{2 \pi}{\kappa_5^2}  \frac{1}{\phi_A^{3/\nu}} , \label{eq:s} \\
n_B (T, \mu_B) &= - \frac{1}{\kappa_5^2} \frac{\Phi_2^\textrm{far}}{\phi_A^{3/\nu} \sqrt{f_0^\textrm{far}}}  \label{eq:nB}
\end{align}
with $T = 1/(4 \pi L \phi_A^{1/\nu} \sqrt{f_0^\textrm{far}})$ and 
$\mu_B = 4 \pi \, T \, \Phi_0^\textrm{far}$.
The needed coefficients $f_0^\textrm{far}$, $\Phi_{0, 2}^\textrm{far}$ and $\phi_A$
are obtained by adjusting the asymptotic series expansions near boundary, 
\begin{align}
f      (r \to \infty) &= f_0^\textrm{far} + \cdots, \\
\phi (r \to \infty) &= \phi_A e^{-\nu \alpha(r) } + \phi_B e^{-\Delta_\phi \alpha(r) } + \cdots, \\
\Phi (r \to \infty) &= \Phi_0^\textrm{far} + \Phi_2^\textrm{far} e^{-2 \alpha(r) } + \cdots, 
\end{align}
to numerical solutions of Eqs.~(\ref{eq:A} - \ref{eq:Phi}) with (\ref{eq:A}' - \ref{eq:Phi}') 
together with the relations
$\nu = 4 - \Delta_\phi$, 
$\Delta_\phi := 2 (1 + \sqrt{1 - 3 p_1})$ (cf.\ Eq.~(\ref{eq:V}) for $p_1$).
Typical dependencies are $f_0^\mathrm{far} \to T$, $\Phi_0^\mathrm{far} \to \mu_B$ and
$\Phi_2^\mathrm{far} \to n_B$.
We emphasize that, using the coordinates (\ref{eq:metric}), is numerically advantageous in comparison with
the ones deployed in \cite{Zollner:2023myk,Zollner:2021stb,Zollner:2020nnt} 
since $r$-dependent quantities appear near boundary as exponentials 
(instead of power functions in the radial bulk coordinate $z$, $z \in [0, z_H]$)
which is mainly favorable for the warp factor $A$. 

In practice, we chose initial conditions set by $\phi_0$ and $\Phi_1$ 
such to generate a rectangular grid over the $T$-$\mu_B$ plane. 
The pressure can be obtained either by a line integral with $p(T=0, \mu_B=0) = 0$ 
or by solving Eq.~(\ref{eq:T(muB)})
with proper boundary conditions at $p(T, \mu_B = 0)$ given by lattice QCD data.
That is, $\textrm{d} p = s \textrm{d} T + n_B \textrm{d} \mu_B$, 
$s = \partial p / \partial T$ and $n_B = \partial p / \partial \mu_B$, is exploited.

\section{Density and ressure at FOPT}\label{sect:appC}

\begin{figure}[ht!]
\includegraphics[width=0.44\columnwidth]{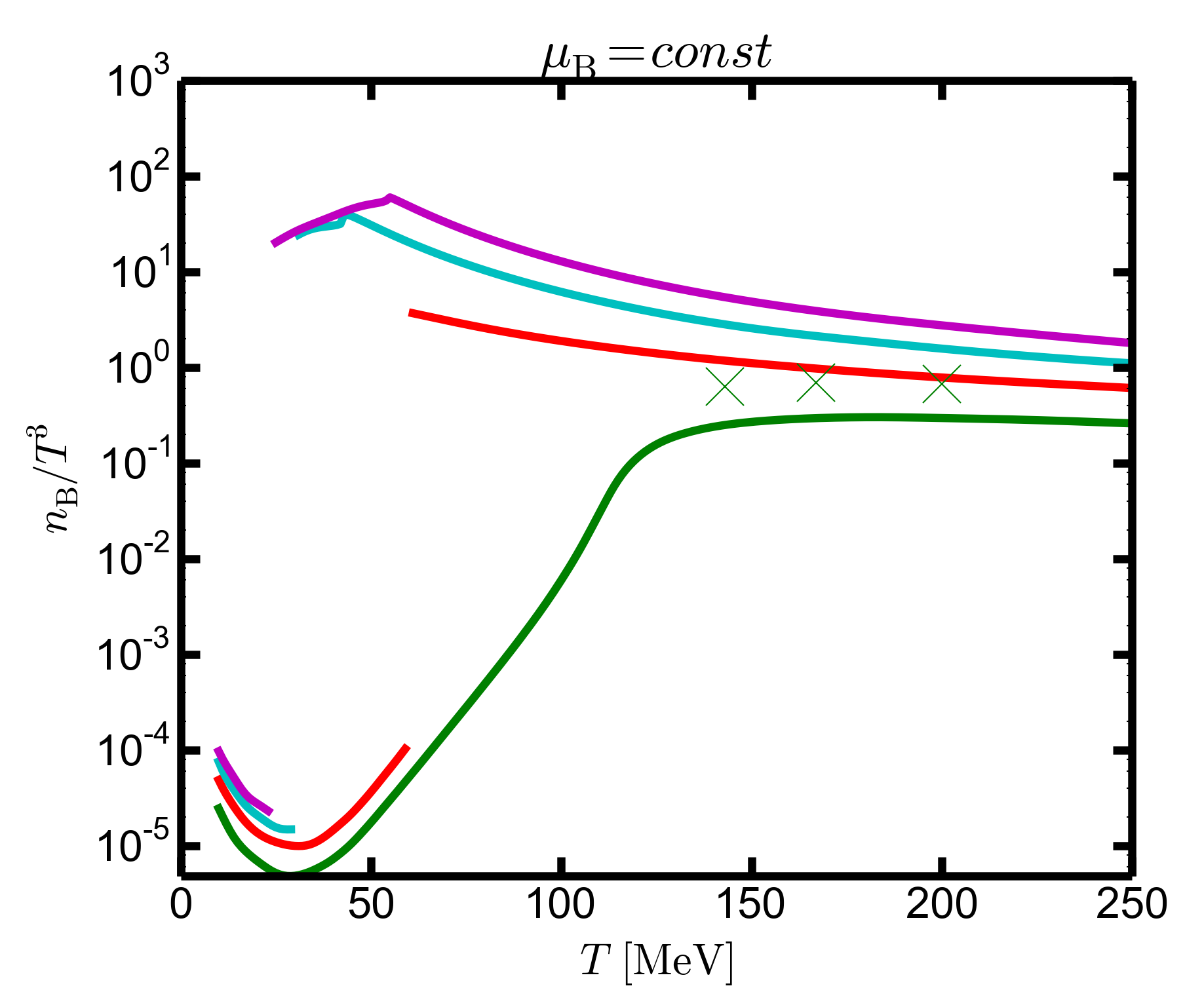}
\includegraphics[width=0.44\columnwidth]{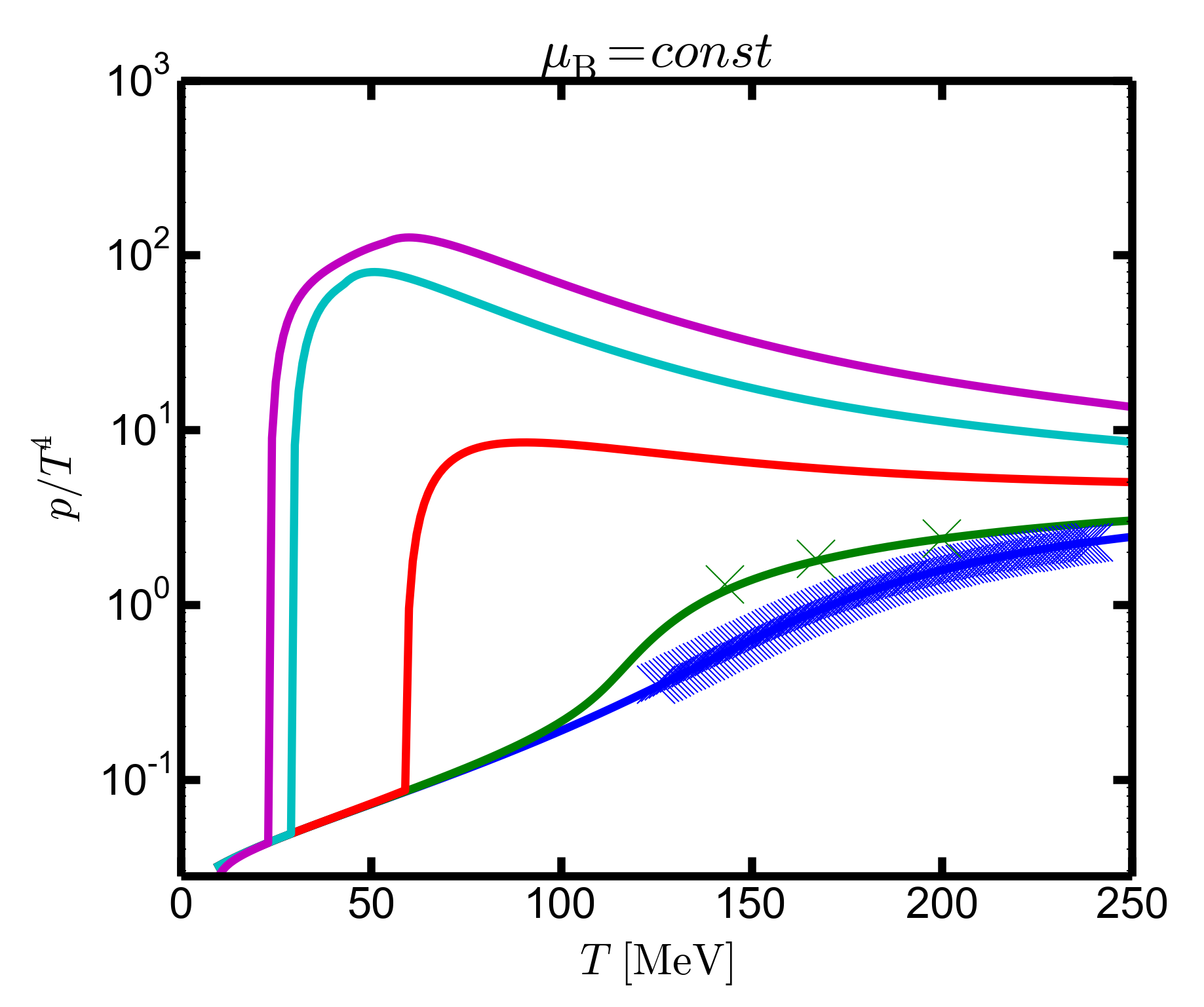}
\caption{The stable branches of 
scaled density $n_B/T^3$ (left panel) and scaled pressure $p/T^4$
as a function of temperature $T$ for various values of 
$\mu_B = n \, 500$~MeV for $n = 0$ (blue), 1 (green), 2 (red), 3 (cyan) and 4 (magenta). 
The crosses depict results of the lattice QCD calculations \cite{Borsanyi:2021sxv}.
\label{fig:p_loop} 
}
\end{figure}

Across the FOPT, the density makes a huge jump, as exhibited in the left panel of  Fig.~\ref{fig:p_loop}.
The FOPT curve is determined by the standard construction: find the self-crossing of the curve
$p/T^4$ as a function of $T$ at $\mu_B = const$.
Above mentioned peculiarities are obvious in the right panel of  Fig.~\ref{fig:p_loop}:
small values of $p/T^4$ even for low temperatures, seemingly $\mu_B$-independence of the low-temperature
branch, and weak $\mu_B$ dependence of the high-temperature branch for $\mu_B \ge 1500$~MeV.   
The squeezing of curves $p = const$ into a narrow corridor right-up to the FOPT curve
(see left panel of Fig.~\ref{fig:contours_p} or right panel of Fig.~\ref{fig:contours_s_nB_p} for $p/T^4$)
is a consequence of the initial steep increase of the high-temperature branch with $T$.

Upon inspecting the right panel of Fig.~\ref{fig:p_loop}, note 
(i) the log scale, 
(ii) the apparent independence of the low-temperature branch on $\mu_B$\footnote{
For some guidance, one could resort to the Hadron Resonance Gas model 
of confined strong-interaction matter with full quantum statistics and vacuum rest masses.
Of course, the pion gas pressure
$p = \frac{3}{2 \pi^2} m_\pi^2 T^2 \sum_{\ell = 1}^\infty \ell^{-2} K_2(\frac{\ell m_\pi}{T})$
is independent of $\mu_B$; it is numerically consistent with the EMd behavior for $T \in [50, 100]$~MeV,
but falls significantly short at $T < 50$~MeV. Adding a nucleon Fermi gas at $\mu_B = 500$~MeV
explains semi-quantitatively the rise of the pressure at $T > 100$~MeV
relative to the $\mu_B = 0$ curve.
All that seems elucidate the failure of such a naive interpretation of the EMd results.
}
(see also left panel in Fig.~\ref{fig:contours_p})
and (iii) the weak $\mu_B$ dependence of the high-temperature branches for
$\mu_B \ge 1500$~MeV which causes the leveling off of the FOPT curve $T_\textrm{FOPT} (\mu_B)$,
see black curves in Figs.~\ref{fig:contours_s_nB_p} and \ref{fig:contours_p}. 

\section{Various dilaton potential parameterizations}\label{sect:appB}

\begin{figure}[t!]
\includegraphics[width=0.30\columnwidth]{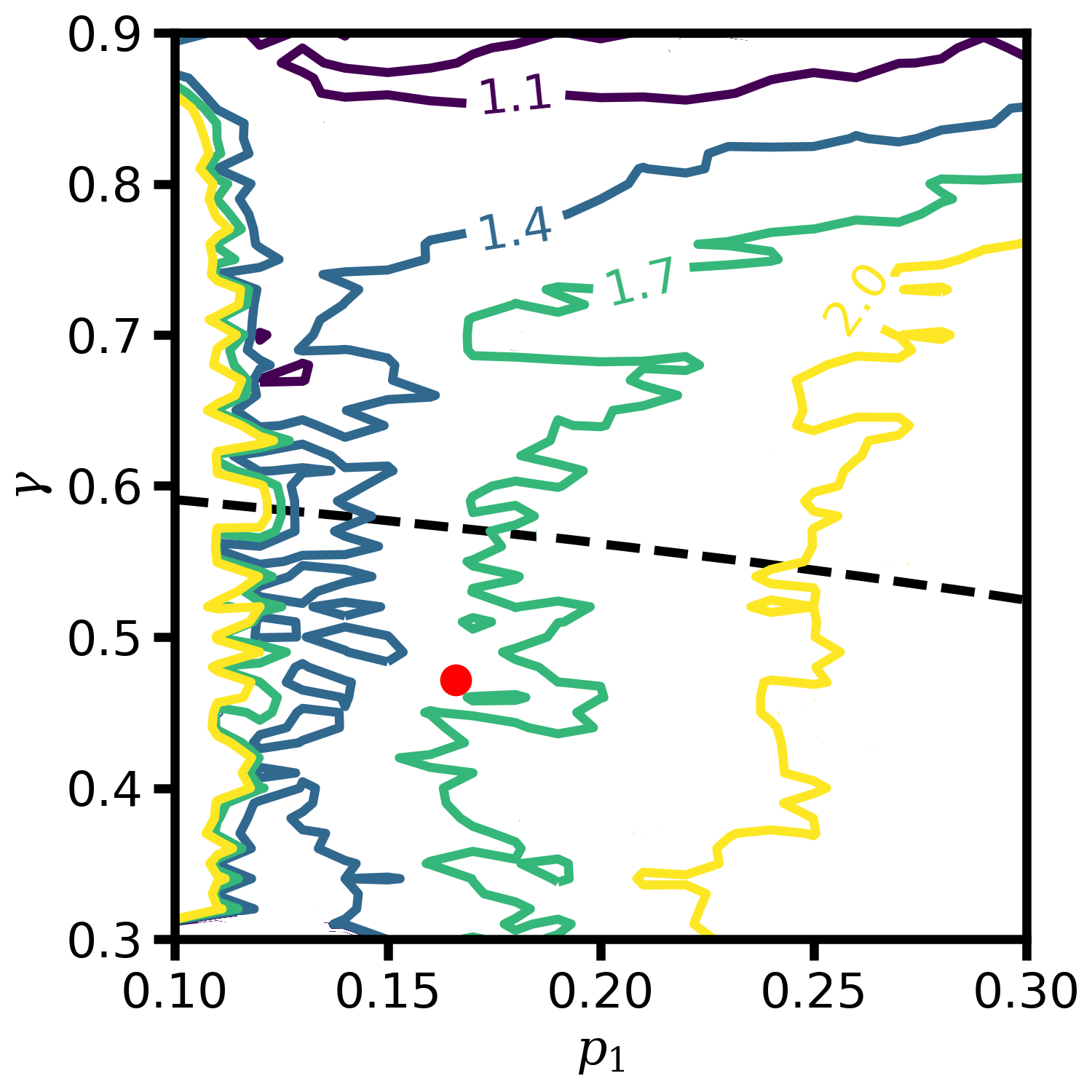} \hspace{3mm}
\includegraphics[width=0.30\columnwidth]{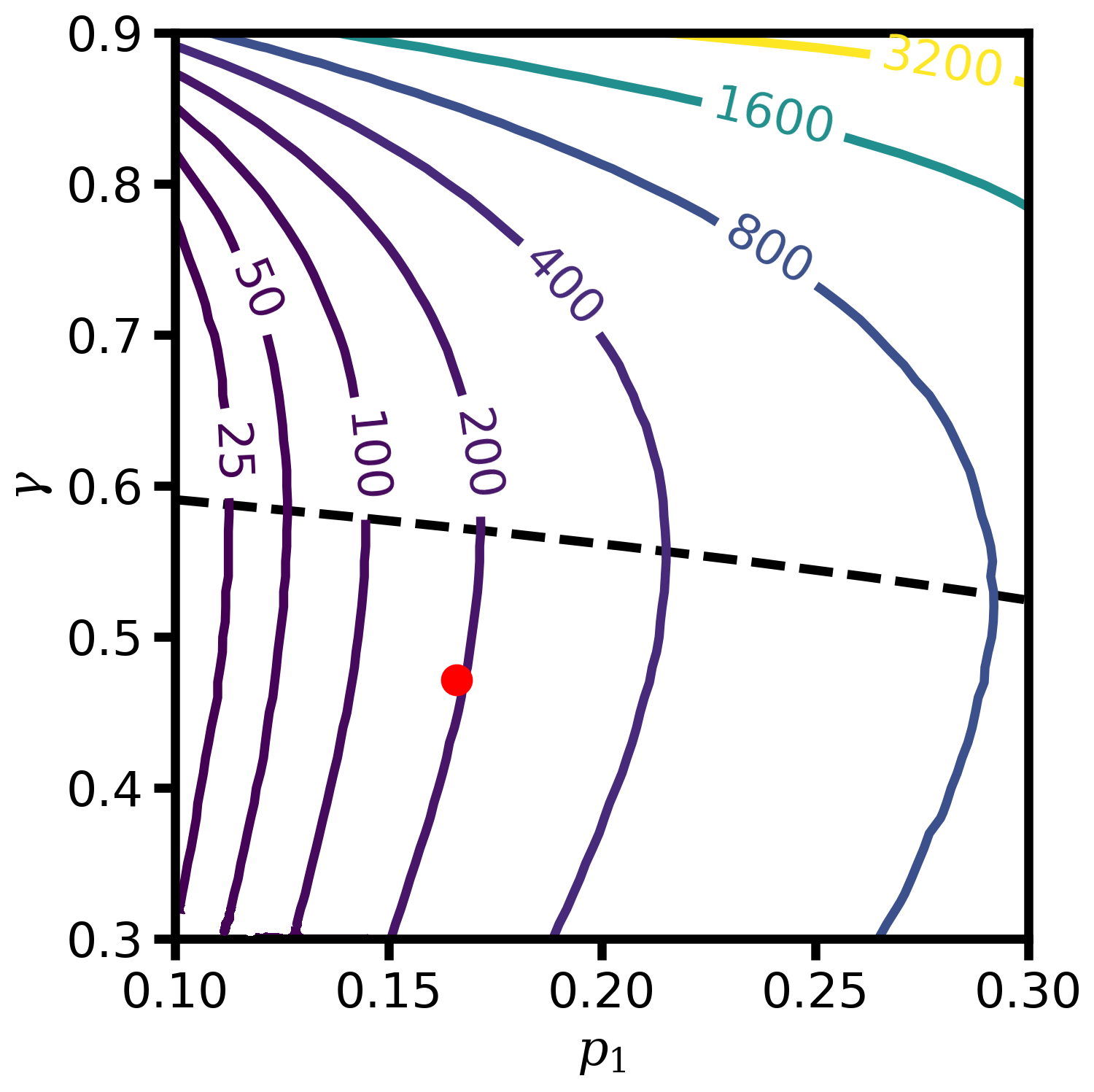} \hspace{3mm}
\includegraphics[width=0.30\columnwidth]{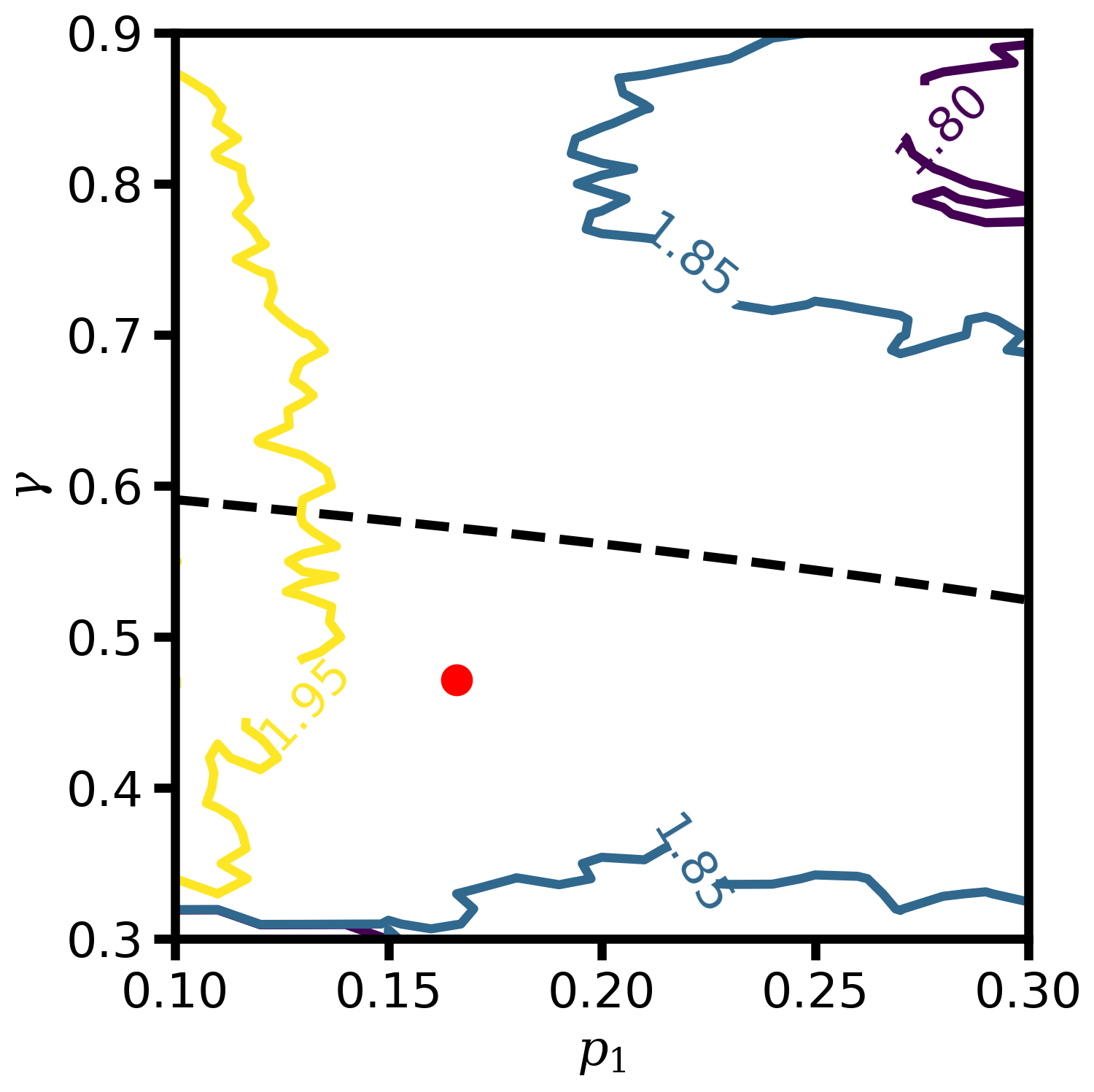}
\caption{Contour plots of $\chi^2$ w.r.t.\ scaled entropy density (left panel), $L^{-1}$ (middle panel) 
and $\kappa_5$ (right panel, in units of $L^{3/2}$)
for the dilaton potential function  Eq.~(\ref{eq:V}) 
with local maximum of $\mathcal{W}_m = 0.6$ at $\phi_m =  3.25$ as side conditions.
The dashed line depicts the locus of $p_3 = 0$ determined by 
$p_1 = \mathcal{W}_m \exp\{ \gamma \phi_m\} (2 - \gamma \phi_m) / \phi_m$,
i.e.\ for $p_3 > 0$, an unintended thermal
phase transition is excluded 
since, beyond the maximum, $\mathcal{W}(\phi)$ is smoothly and monotonously approaching zero at $\phi \to \infty$.
The bullet in the $p_3 < 0$ region is for the parameter choice 
of $p_{1, 2, 3}$ and $\gamma$ listed below Eq.~(\ref{eq:F})
which facilitates $\mathcal{W}_m \approx 0.6$ at $\phi_m \approx  3.25$.
\label{fig:p1_gamma} 
}
\end{figure}

With lacking strict gravity dual of QCD, one adjusts in bottom-up approaches the dilaton potential
at suitable QCD input. General guidelines are described in \cite{Gursoy:2010fj}
and used in \cite{Jarvinen:2021jbd}. Practitioners would prefer to utilize a less theory-based
ansatz and extend the set of parameters to be fixed such to catch numerically the wanted quantities,
as done in Eq.~(\ref{eq:V}). Another dilaton potential ansatz is proposed in \cite{Knaute:2017opk};
it differs from ans\"atze in \cite{DeWolfe:2010he,Grefa:2021qvt,Critelli:2017oub}.
Inspecting plots (not displayed)
of the dilaton potential function $\mathcal{W}(\phi) = \mathrm{d} \log V (\phi) / \mathrm{d} \phi$, 
one recognizes that  
\cite{Knaute:2017opk,Grefa:2021qvt,Critelli:2017oub,Cai:2022omk} look strikingly the same
for $\phi < 3.5$: they display a maximum of $\mathcal{W}_m \approx 0.6$ at $\phi_m \approx 3.25$.
In adiabatic approximation, the maximum causes the minimum of squared sound velocity,
$v_s^2 \approx \frac13 - \frac12 \mathcal{W}$.
The mentioned dilaton potentials can be well fitted by our ansatz Eq.~(\ref{eq:V}) with some spread of
the coefficients $p_{1,2,3,}$ and $\gamma$. The two side conditions make $\mathcal{W}(\phi)$ 
dependent on $p_1$ (which fixes the dynamical dimension $\Delta_\phi$) and $\gamma$:
$p_2 = (-2 p_1 \phi_m +\mathcal{W}_m \exp\{\gamma \phi_m \} [3 - \gamma \phi_m])/\phi_m^2$ and
$p_3 =  (p_1 \phi_m -\mathcal{W}_m \exp\{\gamma \phi_m \} [2 - \gamma \phi_m])/\phi_m^3$. 
The corresponding contour plots of $\chi^2$ w.r.t.\ scaled entropy density, $L{_1}$ and $\kappa_5$
are exhibited in Fig.~\ref{fig:p1_gamma}.
We define $\chi^2 = \sum_{n=1}^{24} 
(s/T^3\vert_{n \, lattice} - s/T^3\vert_{T_n \, \mathcal{W}(p_1,\gamma)})^2
/ (\delta s/T^3\vert_{n \, lattice})^2$ with symmetrized error bars $\delta s/T^3\vert_{n \, lattice}$
at the $24$ values $T_n$ of lattice data \cite{Borsanyi:2021sxv}.
At each point $(p_1, \gamma)$, $\chi^2$ is minimized by free and independent variations of
$L$ and $\kappa_5$.
Surprising is the wide variation of the scale setting parameter $L^{-1}$,
while $\kappa_5/L^{3/2} \approx 1.87$. 
The flickering $\chi^2$ contours are understood
as the result of biasing the fit problem by fixing the maximum of $\mathcal{W}$ at $\phi_m = 3.25$
which flattens the $\chi^2$ landscape. 
The left panel in Fig.~\ref{fig:p1_gamma} suggests that our choice of ansatz and parameters in 
Eq.~(\ref{eq:V}) is not the optimum one when prescribing $\mathcal{W}_m = 0.6$ at $\phi_m = 3.25$.
However, the real $\chi^2$ minimum requires a scan throught the full
$(p_{1, 2, 3}, \gamma, L, \kappa_5)$ space without constraints,
preferentially including other observables too.

The dilaton parameterizations proposed in \cite{Jokela:2024xgz,Hippert:2023bel},
which also reproduce nicely the lattice QCD data,
can not be described quantitatively by our ansatz Eq.~(\ref{eq:V}) in the range $\phi = 0$ up to and including
the first local maximum of $\mathcal{W}$.

\end{appendix}

\begin{acknowledgements}

The authors are grateful to Sz.~Bors\'{a}nyi for communications w.r.t.\ lattice QCD data. 
One author (BK) acknowledges conversations with J.~Erdmenger, K.~Redlich and W.~Weise.
The work is supported in part by the European Union’s Horizon 2020 research
and innovation program STRONG-2020 under grant agreement No 824093. 

\end{acknowledgements}

{}


\begin{thebibliography}{99}

\bibitem{LIGOScientific:2020aai}
B.~P.~Abbott \textit{et al.} [LIGO Scientific and Virgo],
``GW190425: Observation of a Compact Binary Coalescence with Total Mass $\sim 3.4 M_{\odot}$,''
Astrophys. J. Lett. \textbf{892}, no.1, L3 (2020)
\arx[b]{2001.01761}{astro-ph.HE}

\bibitem{Miller:2019cac}
M.~C.~Miller, F.~K.~Lamb, A.~J.~Dittmann, S.~Bogdanov, Z.~Arzoumanian, K.~C.~Gendreau, S.~Guillot, A.~K.~Harding, W.~C.~G.~Ho and J.~M.~Lattimer, \textit{et al.}
``PSR J0030+0451 Mass and Radius from $NICER$ Data and Implications for the Properties of Neutron Star Matter,''
Astrophys. J. Lett. \textbf{887}, no.1, L24 (2019)
\arx[b]{1912.05705}{astro-ph.HE}

\bibitem{Riley:2019yda}
T.~E.~Riley, A.~L.~Watts, S.~Bogdanov, P.~S.~Ray, R.~M.~Ludlam, S.~Guillot, Z.~Arzoumanian, C.~L.~Baker, A.~V.~Bilous and D.~Chakrabarty, \textit{et al.}
``A $NICER$ View of PSR J0030+0451: Millisecond Pulsar Parameter Estimation,''
Astrophys. J. Lett. \textbf{887}, no.1, L21 (2019)
\arx[b]{1912.05702}{astro-ph.HE}

\bibitem{Miller:2021qha}
M.~C.~Miller, F.~K.~Lamb, A.~J.~Dittmann, S.~Bogdanov, Z.~Arzoumanian, K.~C.~Gendreau, S.~Guillot, W.~C.~G.~Ho, J.~M.~Lattimer and M.~Loewenstein, \textit{et al.}
``The Radius of PSR J0740+6620 from NICER and XMM-Newton Data,''
Astrophys. J. Lett. \textbf{918}, no.2, L28 (2021)
\arx[b]{2105.06979}{astro-ph.HE}

\bibitem{Jarvinen:2021jbd}
M.~J\"arvinen,
``Holographic modeling of nuclear matter and neutron stars,''
Eur. Phys. J. C \textbf{82}, no.4, 282 (2022)
\arx[b]{2110.08281}{hep-ph}

\bibitem{Hoyos:2021uff}
C.~Hoyos, N.~Jokela and A.~Vuorinen,
``Holographic approach to compact stars and their binary mergers,''
Prog. Part. Nucl. Phys. \textbf{126}, 103972 (2022)
\arx[b]{2112.08422}{hep-th}

\bibitem{Chesler:2019osn}
P.~M.~Chesler, N.~Jokela, A.~Loeb and A.~Vuorinen,
``Finite-temperature Equations of State for Neutron Star Mergers,''
Phys. Rev. D \textbf{100}, no.6, 066027 (2019)
\arx[b]{1906.08440}{astro-ph.HE}

\bibitem{DeWolfe:2010he}
O.~DeWolfe, S.~S.~Gubser and C.~Rosen,
``A holographic critical point,''
Phys. Rev. D \textbf{83}, 086005 (2011)
\arx[b]{1012.1864}{hep-th}

\bibitem{DeWolfe:2011ts}
O.~DeWolfe, S.~S.~Gubser and C.~Rosen,
``Dynamic critical phenomena at a holographic critical point,''
Phys. Rev. D \textbf{84}, 126014 (2011)
\arx[b]{1108.2029}{hep-th}

\bibitem{Cai:2022omk}
R.~G.~Cai, S.~He, L.~Li and Y.~X.~Wang,
``Probing QCD critical point and induced gravitational wave by black hole physics,''
Phys. Rev. D \textbf{106}, no.12, L121902 (2022)
\arx[b]{2201.02004}{hep-th}

\bibitem{Grefa:2021qvt}
J.~Grefa, J.~Noronha, J.~Noronha-Hostler, I.~Portillo, C.~Ratti and R.~Rougemont,
``Hot and dense quark-gluon plasma thermodynamics from holographic black holes,''
Phys. Rev. D \textbf{104}, no.3, 034002 (2021)
\arx[b]{2102.12042}{nucl-th}

\bibitem{Critelli:2017oub}
R.~Critelli, J.~Noronha, J.~Noronha-Hostler, I.~Portillo, C.~Ratti and R.~Rougemont,
``Critical point in the phase diagram of primordial quark-gluon matter from black hole physics,''
Phys. Rev. D \textbf{96}, no.9, 096026 (2017)
\arx[b]{1706.00455}{nucl-th}

\bibitem{Knaute:2017opk}
J.~Knaute, R.~Yaresko and B.~K\"ampfer,
Phys. Lett. B \textbf{778}, 419-425 (2018)
\arx[b]{1702.06731}{hep-ph}

\bibitem{Grefa:2022sav}
J.~Grefa, M.~Hippert, J.~Noronha, J.~Noronha-Hostler, I.~Portillo, C.~Ratti and R.~Rougemont,
``Transport coefficients of the quark-gluon plasma at the critical point and across the first-order line,''
Phys. Rev. D \textbf{106}, no.3, 034024 (2022)
\arx[b]{2203.00139}{nucl-th}

\bibitem{Zhang:2022uin}
L.~Zhang and M.~Huang,
``Holographic cold dense matter constrained by neutron stars,''
Phys. Rev. D \textbf{106}, no.9, 096028 (2022)
\arx[b]{2209.00766}{nucl-th}

\bibitem{Rougemont:2023gfz}
R.~Rougemont, J.~Grefa, M.~Hippert, J.~Noronha, J.~Noronha-Hostler, I.~Portillo and C.~Ratti,
``Hot QCD phase diagram from holographic Einstein\textendash{}Maxwell\textendash{}Dilaton models,''
Prog. Part. Nucl. Phys. \textbf{135}, 104093 (2024)
\arx[b]{2307.03885}{nucl-th}

\bibitem{Jokela:2024xgz}
N.~Jokela, M.~J\"arvinen and A.~Piispa,
``Refining holographic models of the quark-gluon plasma,''
\arx[b]{2405.02394}{hep-th}

\bibitem{Borsanyi:2021sxv}
S.~Bors\'anyi, Z.~Fodor, J.~N.~Guenther, R.~Kara, S.~D.~Katz, P.~Parotto, A.~P\'asztor, C.~Ratti and K.~K.~Szab\'o,
``Lattice QCD equation of state at finite chemical potential from an alternative expansion scheme,''
Phys. Rev. Lett. \textbf{126}, no.23, 232001 (2021)
\arx[b]{2102.06660}{hep-lat}

\bibitem{HotQCD:2014kol}
A.~Bazavov \textit{et al.} [HotQCD],
``Equation of state in ( 2+1 )-flavor QCD,''
Phys. Rev. D \textbf{90}, 094503 (2014)
\arx[b]{1407.6387}{hep-lat}

\bibitem{Borsanyi:2013bia}
S.~Borsanyi, Z.~Fodor, C.~Hoelbling, S.~D.~Katz, S.~Krieg and K.~K.~Szabo,
``Full result for the QCD equation of state with 2+1 flavors,''
Phys. Lett. B \textbf{730}, 99-104 (2014)
\arx[b]{1309.5258}{hep-lat}

\bibitem{Bazavov:2017dus}
A.~Bazavov, H.~T.~Ding, P.~Hegde, O.~Kaczmarek, F.~Karsch, E.~Laermann, Y.~Maezawa, S.~Mukherjee, H.~Ohno and P.~Petreczky, \textit{et al.}
``The QCD Equation of State to $\mathcal{O}(\mu_B^6)$ from Lattice QCD,''
Phys. Rev. D \textbf{95}, no.5, 054504 (2017)
\arx[b]{1701.04325}{hep-lat}

\bibitem{Borsanyi:2022qlh}
S.~Borsanyi, J.~N.~Guenther, R.~Kara, Z.~Fodor, P.~Parotto, A.~Pasztor, C.~Ratti and K.~K.~Szabo,
``Resummed lattice QCD equation of state at finite baryon density: Strangeness neutrality and beyond,''
Phys. Rev. D \textbf{105}, no.11, 114504 (2022)
\arx[b]{2202.05574}{hep-lat}

\bibitem{Bollweg:2022fqq}
D.~Bollweg \textit{et al.} [HotQCD],
``Equation of state and speed of sound of (2+1)-flavor QCD in strangeness-neutral matter at nonvanishing net baryon-number density,''
Phys. Rev. D \textbf{108}, no.1, 014510 (2023)
\arx[b]{2212.09043}{hep-lat}

\bibitem{Ecker:2022dlg}
C.~Ecker and L.~Rezzolla,
``Impact of large-mass constraints on the properties of neutron stars,''
Mon. Not. Roy. Astron. Soc. \textbf{519}, no.2, 2615-2622 (2022)
\arx[b]{2209.08101}{astro-ph.HE}

\bibitem{Annala:2021gom}
E.~Annala, T.~Gorda, E.~Katerini, A.~Kurkela, J.~N\"attil\"a, V.~Paschalidis and A.~Vuorinen,
``Multimessenger Constraints for Ultradense Matter,''
Phys. Rev. X \textbf{12}, no.1, 011058 (2022)
\arx[b]{2105.05132}{astro-ph.HE}

\bibitem{Hebeler:2013nza}
K.~Hebeler, J.~M.~Lattimer, C.~J.~Pethick and A.~Schwenk,
``Equation of state and neutron star properties constrained by nuclear physics and observation,''
Astrophys. J. \textbf{773}, 11 (2013)
\arx[b]{1303.4662}{astro-ph.SR}

\bibitem{Stephanov:1999zu}
M.~A.~Stephanov, K.~Rajagopal and E.~V.~Shuryak,
``Event-by-event fluctuations in heavy ion collisions and the QCD critical point,''
Phys. Rev. D \textbf{60} (1999) 114028
\arx[o]{9903292}{hep-ph}

\bibitem{Karsch:2001cy}
F.~Karsch,
``Lattice QCD at high temperature and density,''
Lect. Notes Phys. \textbf{583} (2002) 209-249
\arx[o]{0106019}{hep-lat}

\bibitem{Fukushima:2010bq}
K.~Fukushima and T.~Hatsuda,
``The phase diagram of dense QCD,''
Rept. Prog. Phys. \textbf{74} (2011) 014001
\arx[b]{1005.4814}{hep-ph}

\bibitem{Halasz:1998qr}
A.~M.~Halasz, A.~D.~Jackson, R.~E.~Shrock, M.~A.~Stephanov and J.~J.~M.~Verbaarschot,
``On the phase diagram of QCD,''
Phys. Rev. D \textbf{58} (1998) 096007
\arx[o]{9804290}{hep-ph}

\bibitem{Almaalol:2022xwv}
D.~Almaalol, M.~Hippert, J.~Noronha-Hostler, J.~Noronha, E.~Speranza, G.~Basar, S.~Bass, D.~Cebra, V.~Dexheimer and D.~Keane, \textit{et al.}
``QCD Phase Structure and Interactions at High Baryon Density: Continuation of BES Physics Program with CBM at FAIR,''
\arx[b]{2209.05009}{nucl-ex}

\bibitem{Du:2024wjm}
L.~Du, A.~Sorensen and M.~Stephanov,
``The QCD phase diagram and Beam Energy Scan physics: a theory overview,''
\arx[b]{2402.10183}{nucl-th}

\bibitem{Lovato:2022vgq}
A.~Lovato, T.~Dore, R.~D.~Pisarski, B.~Schenke, K.~Chatziioannou, J.~S.~Read, P.~Landry, P.~Danielewicz, D.~Lee and S.~Pratt, \textit{et al.}
``Long Range Plan: Dense matter theory for heavy-ion collisions and neutron stars,''
\arx[b]{2211.02224}{nucl-th}

\bibitem{MUSES:2023hyz}
R.~Kumar \textit{et al.} [MUSES],
``Theoretical and Experimental Constraints for the Equation of State of Dense and Hot Matter,''
\arx[b]{2303.17021}{nucl-th}

\bibitem{Fujimoto:2022ohj}
Y.~Fujimoto, K.~Fukushima, L.~D.~McLerran and M.~Praszalowicz,
``Trace Anomaly as Signature of Conformality in Neutron Stars,''
Phys. Rev. Lett. \textbf{129}, no.25, 252702 (2022)
\arx[b]{2207.06753}{nucl-th}

\bibitem{Marczenko:2022jhl}
M.~Marczenko, L.~McLerran, K.~Redlich and C.~Sasaki,
``Reaching percolation and conformal limits in neutron stars,''
Phys. Rev. C \textbf{107}, no.2, 025802 (2023)
\arx[b]{2207.13059}{nucl-th}

\bibitem{Pradeep:2024cca}
M.~S.~Pradeep, N.~Sogabe, M.~Stephanov and H.~U.~Yee,
``Non-monotonic specific entropy on the transition line near the QCD critical point,''
\arx[b]{2402.09519}{nucl-th}

\bibitem{He:2023ado}
S.~He, L.~Li, S.~Wang and S.~J.~Wang,
``Constraints on holographic QCD phase transitions from PTA observations,''
\arx[b]{2308.07257}{hep-ph}

\bibitem{Middeldorf-Wygas:2020glx}
M.~M.~Middeldorf-Wygas, I.~M.~Oldengott, D.~B\"odeker and D.~J.~Schwarz,
``Cosmic QCD transition for large lepton flavor asymmetries,''
Phys. Rev. D \textbf{105}, no.12, 123533 (2022)
\arx[b]{2009.00036}{hep-ph}

\bibitem{Bellwied:2015lba}
R.~Bellwied, S.~Borsanyi, Z.~Fodor, S.~D.~Katz, A.~Pasztor, C.~Ratti and K.~K.~Szabo,
``Fluctuations and correlations in high temperature QCD,''
Phys. Rev. D \textbf{92}, no.11, 114505 (2015)
\arx[b]{1507.04627}{hep-lat}

\bibitem{Yaresko:2015ysa}
R.~Yaresko, J.~Knaute and B.~K\"ampfer,
``Cross-over versus first-order phase transition in holographic gravity\textendash{}single-dilaton models of QCD thermodynamics,''
Eur. Phys. J. C \textbf{75}, no.6, 295 (2015)
\arx[b]{1503.09065}{hep-ph}

\bibitem{Zollner:2018uep}
R.~Z\"ollner and B.~K\"ampfer,
``Phase structures emerging from holography with Einstein gravity -- dilaton models at finite temperature,''
Eur. Phys. J. Plus \textbf{135}, no.3, 304 (2020)
\arx[b]{1807.04260}{hep-th}

\bibitem{Hippert:2023bel}
M.~Hippert, J.~Grefa, T.~A.~Manning, J.~Noronha, J.~Noronha-Hostler, I.~Portillo Vazquez, C.~Ratti, R.~Rougemont and M.~Trujillo,
``Bayesian location of the QCD critical point from a holographic perspective,''
\arx[b]{2309.00579}{nucl-th}

\bibitem{Zollner:2023myk}
R.~Z\"ollner, M.~Ding and B.~K\"ampfer,
``Masses of compact (neutron) stars with distinguished cores,''
Particles \textbf{6}, no.1, 217-238 (2023)
\arx[b]{2302.01389}{nucl-th}

\bibitem{Gao:2020fbl}
F.~Gao and J.~M.~Pawlowski,
``Chiral phase structure and critical end point in QCD,''
Phys. Lett. B \textbf{820}, 136584 (2021)
\arx[b]{2010.13705}{hep-ph}

\bibitem{Bernhardt:2021iql}
J.~Bernhardt, C.~S.~Fischer, P.~Isserstedt and B.~J.~Schaefer,
``Critical endpoint of QCD in a finite volume,''
Phys. Rev. D \textbf{104}, no.7, 074035 (2021)
\arx[b]{2107.05504}{hep-ph}

\bibitem{Karthein:2021nxe}
J.~M.~Karthein, D.~Mroczek, A.~R.~Nava Acuna, J.~Noronha-Hostler, P.~Parotto, D.~R.~P.~Price and C.~Ratti,
``Strangeness-neutral equation of state for QCD with a critical point,''
Eur. Phys. J. Plus \textbf{136}, no.6, 621 (2021)
\arx[b]{2103.08146}{hep-ph}

\bibitem{Parotto:2018pwx}
P.~Parotto, M.~Bluhm, D.~Mroczek, M.~Nahrgang, J.~Noronha-Hostler, K.~Rajagopal, C.~Ratti, T.~Sch\"afer and M.~Stephanov,
``QCD equation of state matched to lattice data and exhibiting a critical point singularity,''
Phys. Rev. C \textbf{101}, no.3, 034901 (2020)
\arx[b]{1805.05249}{hep-ph}

\bibitem{Flor:2020fdw}
F.~A.~Flor, G.~Olinger and R.~Bellwied,
``Flavour and Energy Dependence of Chemical Freeze-out Temperatures in Relativistic Heavy Ion Collisions from RHIC-BES to LHC Energies,''
Phys. Lett. B \textbf{814}, 136098 (2021)
\arx[b]{2009.14781}{nucl-ex}

\bibitem{Gunther:2017sxn}
J.~G\"unther, R.~Bellwied, S.~Borsanyi, Z.~Fodor, S.~D.~Katz, A.~Pasztor and C.~Ratti,
``The QCD equation of state at finite density from analytical continuation,''
EPJ Web Conf. \textbf{137}, 07008 (2017)
\arX{10.1051/epjconf/201713707008}{}

\bibitem{Zollner:2020nnt}
R.~Z\"ollner and B.~K\"ampfer,
``Quarkonia Formation in a Holographic Gravity\textendash{}Dilaton Background Describing QCD Thermodynamics,''
Particles \textbf{4}, no.2, 159-177 (2021)
\arx[b]{2007.14287}{hep-ph}

\bibitem{Zollner:2021stb}
R.~Z\"ollner and B.~K\"ampfer,
``Holographic bottomonium formation in a cooling strong-interaction medium at finite baryon density,''
Phys. Rev. D \textbf{104}, no.10, 106005 (2021)
\arx[b]{2109.05824}{hep-th}

\bibitem{Gursoy:2010fj}
U.~Gursoy, E.~Kiritsis, L.~Mazzanti, G.~Michalogiorgakis and F.~Nitti,
``Improved Holographic QCD,''
Lect. Notes Phys. \textbf{828}, 79-146 (2011)
\arx[b]{1006.5461}{hep-th}


\end{thebibliography}
\end{document}